\documentclass[conference]{IEEEtran}

\usepackage{cite}
\usepackage{subfigure}
\usepackage{amsmath}
\usepackage{tabularx}
\usepackage{color}
\usepackage{xspace}
\usepackage{graphicx}    % For importing graphics
\usepackage[algoruled,resetcount,linesnumbered]{algorithm2e}
\usepackage{capt-of}
\usepackage{multirow}
\usepackage{csquotes} % for \enquote{}
\usepackage{amsfonts} % for \mathbb{}
\usepackage{mathrsfs}
\usepackage{bm}
\usepackage{url} 
\usepackage{balance}
\usepackage{cleveref} % fro \cref
\usepackage{amsthm} % http://ctan.org/pkg/amsthm

\usepackage{gensymb}  % for degree symbol
\usepackage[utf8x]{inputenc}
\usepackage{enumitem,kantlipsum}
\SetAlFnt{\small} % set font size in Algorithm
\SetKwInput{KwInput}{Input}                % Set the Input
\SetKwInput{KwOutput}{Output}              % set the Output
\usepackage[T1]{fontenc}

\begin{document}
\title{Communication-Efficient Federated Learning for Heterogeneous Edge Devices Based on Adaptive Gradient Quantization}

\author{\IEEEauthorblockN{Heting Liu, Fang He and Guohong Cao}
\IEEEauthorblockA{Department of Computer Science and Engineering \\
The Pennsylvania State University\\
Email: \{hxl476, fxh35, gxc27\}@psu.edu}
% \and
% \IEEEauthorblockN{Fang He}
% \IEEEauthorblockA{\textit{Department of Computer Science and Engineering} \\
% \textit{The Pennsylvania State University}\\
% fxh35@psu.edu}
% \and
% \IEEEauthorblockN{Guohong Cao}
% \IEEEauthorblockA{\textit{Department of Computer Science and Engineering} \\
% \textit{The Pennsylvania State University}\\
% gxc27@psu.edu}
}

\maketitle
\begin{abstract}
Federated learning (FL) enables geographically dispersed edge devices (i.e., clients) to learn a global model without sharing the local datasets, where each client performs gradient descent with its local data and uploads the gradients to a central server to update the global model.
However, FL faces massive communication overhead resulted from uploading the gradients in each training round. 
To address this problem,  most existing research compresses the gradients with fixed and unified quantization for all the clients, which neither seeks adaptive quantization due to the varying gradient norms at different rounds, nor exploits the heterogeneity of the clients to accelerate FL.
In this paper, we propose a novel adaptive and heterogeneous gradient quantization algorithm (AdaGQ) for FL to minimize the wall-clock training time from two aspects: i) adaptive quantization which exploits the change of gradient norm to adjust the quantization resolution in each training round; and ii) heterogeneous quantization which assigns lower quantization resolution to slow clients to align their training time with other clients to mitigate the communication bottleneck, and higher quantization resolution to fast clients to achieve a better communication efficiency and accuracy tradeoff.
Evaluations based on various models and datasets validate the benefits of AdaGQ, reducing the total training time by up to 52.1\% compared to baseline algorithms (e.g., FedAvg, QSGD).
\end{abstract}

\IEEEpeerreviewmaketitle
%------------------------------------------------------------------------
%********** Introduction ***************
\section{Introduction}
\label{sec:intro}

Intelligent applications based on deep neural networks (DNN) have been developed for edge devices such as Internet of things (IoTs) and smart mobile devices over the past years \cite{abbas-2017survey, liu-twc2022}. 
These applications rely heavily on 
the knowledge obtained from the big data, and they generate massive amounts of data in return.
The most straightforward way to utilize these locally generated data is to upload the data to the cloud and train the DNN models in the cloud \cite{lim-2020federated-survey, liu-tvt2021}. However, sharing data is challenging due to the increasing privacy concerns. 

Federated learning (FL) \cite{konevcny-arxiv2016, bonawitz-mlsys2019} emerges as a solution to privacy-preserving machine learning. 
In FL, multiple devices train a shared global model without uploading their data to the central server.
Specifically, in each round every participating device (i.e., client) does the following. 
First, it downloads the latest model from the central server. Next, it updates the downloaded model based on its local data using stochastic gradient descent (SGD). Finally, all clients upload their model updates to the central server, where the 
model updates are aggregated to form a new global model. 
These steps are repeated until a certain convergence criterion is satisfied.

One important research problem in FL is to address the massive communication overhead resulted from uploading the gradients and downloading in each training round. The model updates (i.e., gradients) %in each uploading/downloading have similar data size as the model itself, which 
can be in the range from megabytes to gigabytes for modern DNN architectures with millions of parameters \cite{chen2021-twc}.  
Thus, communication will become a bottleneck when applying FL to edge devices 
where the wireless bandwidth is limited.
Existing approaches addressing the communication overhead in FL fall into two folds:
i) to reduce the amount of communication by allowing each client to perform multiple
local updates between two communication (aggregation) rounds \cite{mcmahan-aistat2017, haddadpour-nips2019}. 
While the number of communication rounds is reduced, the data size in each communication round is still very high; and ii) to mitigate the communication overhead by using gradient compression \cite{aji-emnlp2017, wen-nips2017, alistarh-nips2017}, reducing the amount of data transmitted in each training round.

One widely used gradient compression method is gradient quantization, where the gradient is represented by a number of bits which can determine the number of quantization levels and affect the performance of the gradient quantization algorithms. 
With fewer number of quantization levels (i.e., low quantization resolution) used, the quantization algorithm uses fewer number of bits and then reduces the communication overhead more aggressively. 
However, it also introduces quantization error in the uploaded gradients, and thus may require more training 
rounds to converge.
With higher quantization resolution, there will be less quantization error, but more data has to be transmitted in each round, increasing the accumulated training time. 
Thus, the quantization resolution should be carefully determined to minimize the wall-clock training time.

Existing gradient quantization algorithms \cite{alistarh-nips2017, reisizadeh-aistats2020, wen-nips2017} rely on fixed and pre-determined quantization throughout the training process. However, different FL task has different characteristics in terms of convergence time, communication cost and network condition, etc., and then 
it makes pre-determined quantization less effective because the optimal quantization resolution at different time 
may be different. For example, based on our measurements, the gradient value has large variations during the training process, and thus we should adaptively adjust the quantization based on the training rounds. 

Moreover, in mobile edge computing, different edge devices have different communication resources and some of them only have limited wireless bandwidth. 
Such heterogeneity makes fixed quantization less effective because the training speed is bounded by the slowest client, thus leading to long waiting time for other clients (i.e., straggler effects).
Recent work \cite{li-mobicom2021, wang-tsp2021, diao-iclr2020, zhao-icdcs2019, ma-2021ijcai} studied FL under heterogeneous clients. For example, \cite{zhao-icdcs2019, ma-2021ijcai} addressed the straggler problem by designing an asynchronous aggregation strategy where clients do not wait for each other every round and simply run independently so that the waiting time of faster clients is reduced.  
Although asynchronous aggregation can reduce the communication time, delaying gradients of the stragglers may introduce 
errors, increase the number of training rounds or even cause divergence in model training.
We take a different approach to address this problem by assigning fewer number of quantization levels to the slow clients. 
In this way, the slowest client will transmit less amount of data in each round, and its transmission time can be reduced and aligned with other clients, thus reducing the overall per-round training time.

To realize our ideas, in this paper, we propose an \underline{Ada}ptive and Heterogeneous \underline{G}radient \underline{Q}uantization algorithm, namely {\em AdaGQ}, which dynamically assigns different number of quantization levels to different clients based on online learning to minimize the wall-clock training time of FL. 
Specifically, the proposed strategy includes two aspects. 
(i) The number of quantization levels should be adaptive to the training process in accordance with the gradient norm. 
Since the gradient norm (indicating the upper bound of the gradient magnitude) has large variations as the training proceeds, 
different numbers of quantization levels are chosen based on the training round to achieve better tradeoff between communication efficiency and accuracy.
(ii) The number of quantization levels should adapt to the clients' communication capability. 
Specifically, slow clients (i.e., the clients with longer local training and communication time) are assigned fewer number of quantization levels to mitigate the delay of gradient aggregation at the server; while fast clients are assigned more quantization levels to maintain the accuracy achieved by the global model. 

This paper has the following main contributions.
\begin{itemize}
    \item Through extensive experiments, we identify that gradient quantization should be adaptive to the training process and the clients' communication capability to reduce the training time for heterogeneous clients.  
    
    \item We design AdaGQ, an online learning based adaptive and heterogeneous gradient quantization, to minimize the wall-clock training time.
    
    \item We evaluate the proposed scheme through extensive experiments with various datasets and deep learning models. Evaluation results show that AdaGQ reduces the total training time by 34.8\%-52.1\% compared to baselines.
\end{itemize}

%********** Background ***************
\section{Background and Motivation}
\label{sec:background}

% *** ML training
In FL, the goal of the training process is to find the model parameters (weights) $\mathbf{w}$ that can minimize a loss function $L(\mathbf{w}) := \frac{1}{D} \sum_{h=1}^D l_h(\mathbf{w})$, where $l_h(\mathbf{w})$ is the loss of data sample $h$ and $D$ is the number of data samples.
In particular, we minimize $L(\mathbf{w})$ using SGD algorithm, i.e., $\mathbf{w}_{k+1} = \mathbf{w}_{k} - \eta_k g(\mathbf{w}_{k})$ for $k\in \{0, 1, \cdots\}$, 
where $g(\mathbf{w}_{k})$ denotes the stochastic gradients at iteration $k$, and $\eta_{k}$ is the step size at iteration $k$.
% *** intro of FL
In FL, the training data is spread across a number of edge devices, and
FL enables distributed training without sharing data across these clients.
Assume there are $n$ clients and a central aggregating server. 
Each client $i \in \{1, \cdots, n\}$ has a dataset $D_i$ of size $m_i$. 
In a typical FL algorithm \cite{haddadpour-nips2019, reisizadeh-aistats2020, wang-infocom2020},
the goal is to train a global model, represented by the parameter vector $\mathbf{w}$,
which minimizes
\begin{equation}
\label{eq:fl_obj}
    \min_{\mathbf{w} \in \mathbb{R}^d} L(\mathbf{w}) = \sum_{i=1}^n p_i L_i(\mathbf{w}),
\end{equation}
where $L_i(\mathbf{w})$ is the loss function at client $i$, and $p_i = \frac{m_i}{\sum_{i=1}^n m_i}$ represents the fraction of data stored at client $i$.

\subsection{Federated Learning with Gradient Quantization}
\label{sec:fl_basic}

% *** gradient quantization
In the conventional FedAvg \cite{mcmahan-aistat2017}, every client performs a certain number of gradient descent steps locally at each round, and then uploads the updated weights to the central server followed by a global aggregation.
This procedure is repeated until the training converges.
In quantized SDG \cite{alistarh-nips2017}, the clients upload quantized gradients instead of model weights.
Formally, let $g(\mathbf{w}_{k}^{(i)})$ denote the stochastic gradient on the $i^{th}$ client's dataset $D_i$ at round $k$.  
To reduce the communication cost at each round, every client sends quantized weight updates (gradients) $Q(g(\mathbf{w}_{k}^{(i)}))$ to the server, where $Q(\cdot)$ represents a stochastic quantization function.
Once the server receives the quantized gradients from all clients, the aggregation is performed to update the global model by Eq.~(\ref{eq:quant_aggregation}):
\begin{equation}
\label{eq:quant_aggregation}
    \mathbf{w}_{k+1} = \mathbf{w}_k - \eta_k \sum_{i=1}^n p_i Q(g(\mathbf{w}_{k}^{(i)})).
\end{equation}

% *** QSGD
We adopt the commonly used stochastic uniform quantization function (QSGD) $Q_s(\cdot)$ \cite{wen-nips2017, alistarh-nips2017}, where $s \in \mathbb{N} = \{1, 2, \cdots\}$ is the parameter that determines the compression resolution, i.e., 
the number of quantization levels.
Let $\mathbf{v}$ denote the aligned gradients: $\mathbf{v} = [v_1, \cdots, v_d] \in \mathbb{R}^d$ with $\mathbf{v} \neq \mathbf{0}$. The $j^{th}$ dimension of $\mathbf{v}$, $v_j$,  is quantized to be $Q_s(v_j)$ as follows,
\begin{equation}
\label{eq:qsdg}
    Q_s(v_j) = ||\mathbf{v}||_2 \cdot sign(v_j) \cdot \zeta_j(\mathbf{v},s),
\end{equation}
where $\zeta_j(\mathbf{v},s)$ is a random variable defined as
\begin{equation}\small
    \zeta_j(\mathbf{v},s) = 
    \left\{\begin{array}{cc}
           l/s, & \quad with \; probability \; (1-\frac{|v_j|}{||\mathbf{v}||_2} s+l)\\
          (l+1)/s, &\quad otherwise.
    \end{array}
    \right.
\end{equation}
Here, $0 \leq l < s$ is an integer such that $\frac{|v_j|}{||\mathbf{v}||_2} \in [l/s, (l+1)/s]$.
$Q_s(\mathbf{v})$ is defined to be $\mathbf{0}$ if $\mathbf{v} = \mathbf{0}$.

The idea of QSGD can be explained as follows. A gradient $v_j$ consists of a sign bit and the absolute value $|v_j|$.
To quantize $|v_j| \in [0, ||\mathbf{v}||_2]$, we divide the interval into $s-1$ bins of equal length, with end points $0 = \tau_1 < \tau_2 < \cdots < \tau_s = ||\mathbf{v}||_2$. 
Given $v_j$ that belongs to a bin $[\tau_i, \tau_{i+1})$, the probability is assigned to represent $v_j$ to be $\tau_i$ or $\tau_{i+1}$ based on its relative location inside the bin. 
That is, $\tau_i$ is chosen to represent $v_j$ with probability $p = 1-(v_j - \tau_{i})/(\tau_{i+1} - \tau_{i})$, 
and $\tau_{i+1}$ is chosen with probability $1-p$ (so that we have $\mathbb{E}[Q_s(v_j)] = v_j$). 
Then, $v_j$ is represented by an end point which only needs $\log_2(s)+1$ bits (with the sign bit).
Different from weight quantization which quantizes the model weights, gradient quantization compresses the gradients 
without changing the number of bits to represent the model weights. 

In QSGD based FL, the number of quantization levels (i.e., quantization resolution) is manually pre-defined, and shared by all clients throughout the model training process, which faces two issues: i) the gradient norm $||\mathbf{v}||_2$ may change in different rounds of the training process, leading to a varying interval $[0, ||\mathbf{v}||_2]$. However, the pre-defined quantization fails to automatically adapt to different intervals; ii) clients have heterogeneous communication resources, which creates opportunities to minimize the total training time by assigning different clients with different quantization resolutions.

Next, we will detail our motivation of adaptive gradient quantization
by inspecting the gradient norm during training.
We also explain how to assign different quantization resolutions to heterogeneous clients to minimize the training time.

\subsection{Motivations}
\label{sec:preliminary}

% Figure 2
\begin{figure}[t]
    \subfigure[Gradient norm v.s. round]{
    \centering
        \includegraphics[width=1.54in]{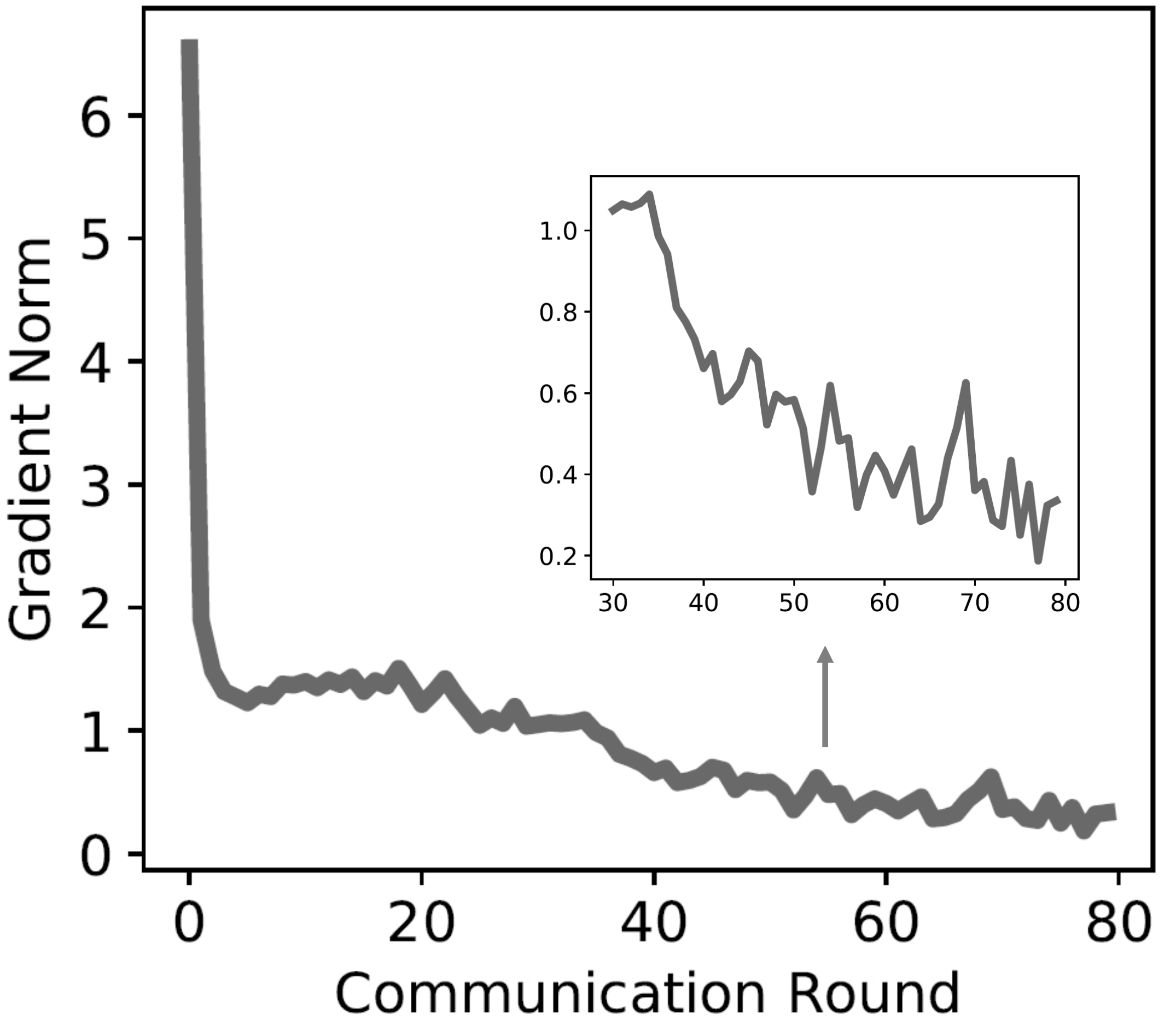}
    }
    \hspace*{-0.8em}
    \subfigure[Accuracy v.s. round]{
    \centering
        \includegraphics[width=1.6in]{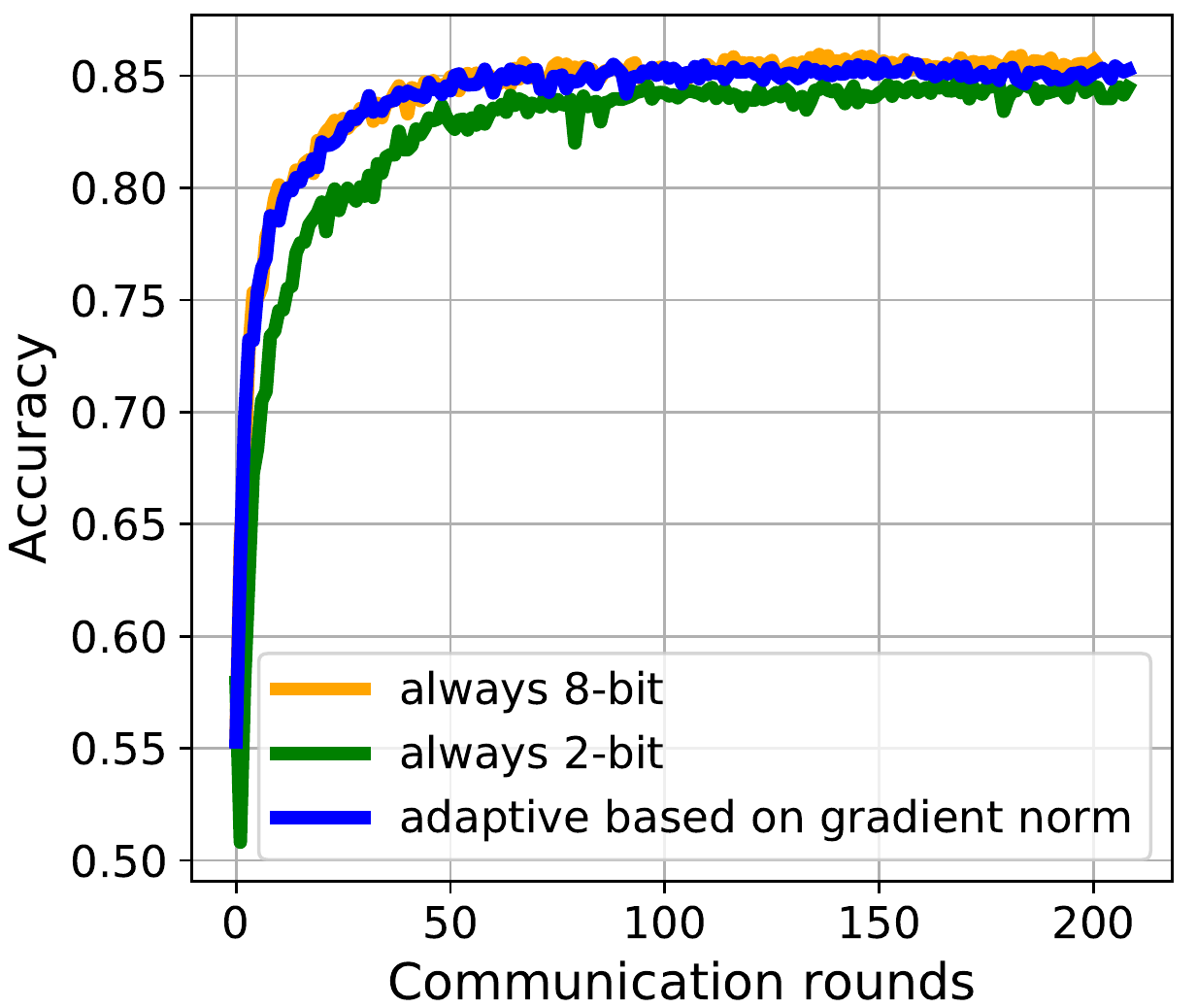}
    }
    \vspace{-0.1in}
    \caption{Training process of ResNet-18 on Cifar-10.}
    \label{fig:gradient_norm}
    \vspace{-0.1in}
\end{figure}

In this section, we investigate the idea of \textit{adaptive gradient quantization} and \textit{heterogeneous gradient quantization}. 
We start with observing the gradient norm during the training of ResNet-18 and GoogLeNet on the Cifar-10 dataset.
As shown in Fig.~\ref{fig:gradient_norm}(a), the gradient norm has large variations in training ResNet-18, 
i.e., with a rapid decrease in the early rounds and mild decrease later on (similar observations for GoogLeNet).
Based on the aforementioned analysis of QSGD, a larger gradient norm, e.g.,  the $||\mathbf{v}||_2$ in the early rounds, 
results in a wide value range of gradients.
Thus, to reduce the quantization error, more quantization levels (i.e., higher quantization resolution) should be used to represent the gradient in the early training rounds. 

On the other hand, a small gradient norm in later training rounds suggests that the gradient has a small value range. 
Then, fewer number of quantization levels will be able to sufficiently represent the gradient with good precision. 
This idea is supported by some other research \cite{gur-2018gradient-arxiv, jastrz-iclr2018}
that highlights the importance of early training phases. 
More importantly, low quantization resolution means less amount of data to be uploaded to the server 
and thus reducing the communication time, potentially reducing the total training time.
The above analysis motivates us to adaptively adjust the quantization resolution based on the gradient norm 
to minimize the training time without compromising accuracy. 
To validate this idea, we adjust the quantization resolution based on the change of gradient norm, i.e., $s_{k} = s_{k-1} + \log \frac{||\mathbf{g_{k}}||}{||\mathbf{g_{k-1}}||}$, where $s_{k}$ denotes the quantization resolution of round $k$ and $||\mathbf{g_{k}}||$ denotes the gradient norm of round $k$, respectively.
Fig.~\ref{fig:gradient_norm}(b) shows the accuracy in each communication round  when training ResNet-18 on Cifar-10 dataset with the above {\em adaptive quantization}. We observe that adaptive quantization achieves similar final accuracy as that achieved by always using 8-bit quantization, higher than that by always using 2-bit quantization. Thus, using lower quantization resolution at later training stage may reduce the total training time (due to less bits transmitted) 
without degrading the performance.
We also have similar observations for GoogLeNet, but not shown due to space limitations. 

% Figure 3.
\begin{figure}[t]
    \subfigure[Accumulated time ]{
    \centering
        \includegraphics[width=1.63in]{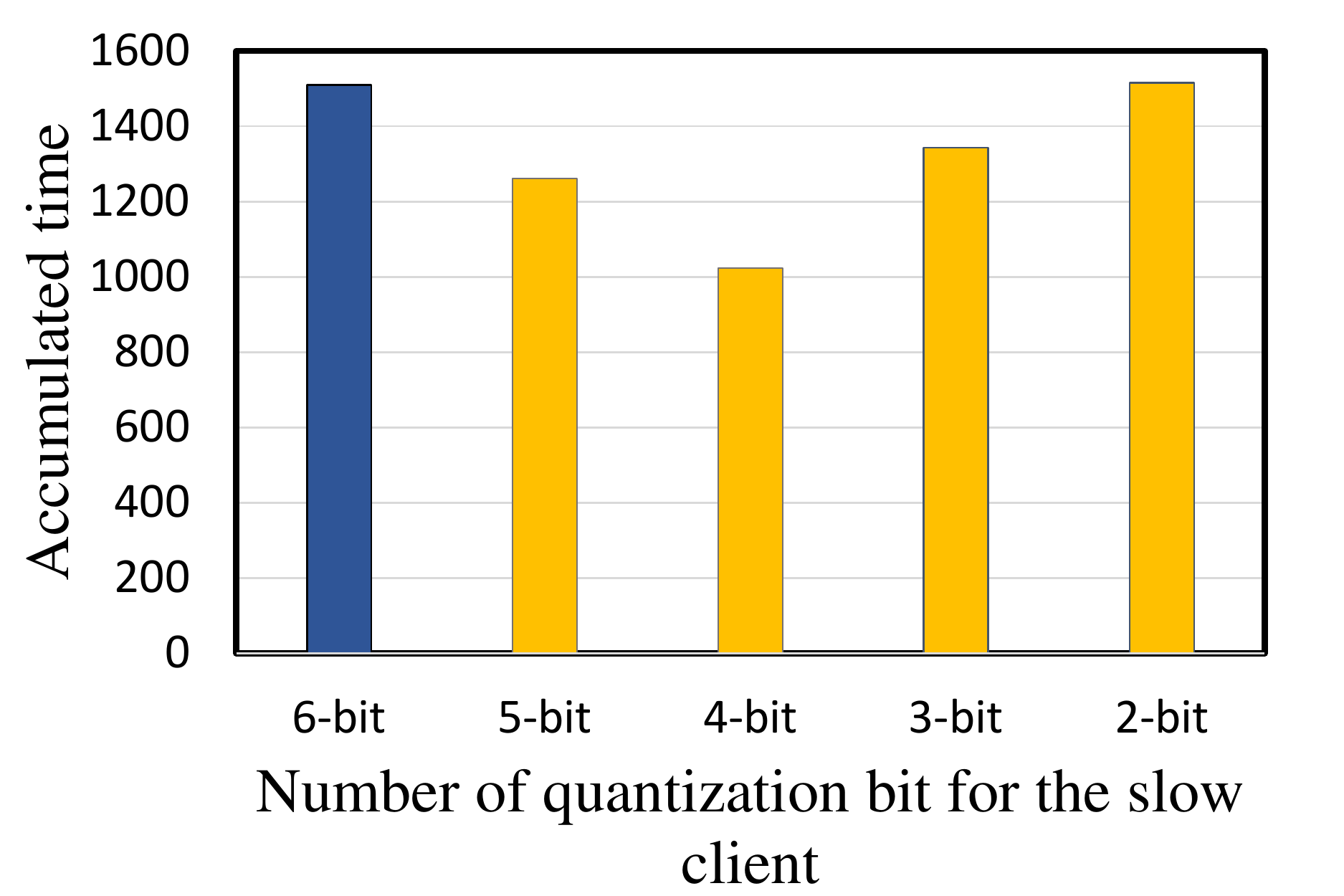}
    }
    \hspace*{-1.5em}
    \subfigure[Accuracy over rounds]{
    \centering
        \includegraphics[width=1.7in]{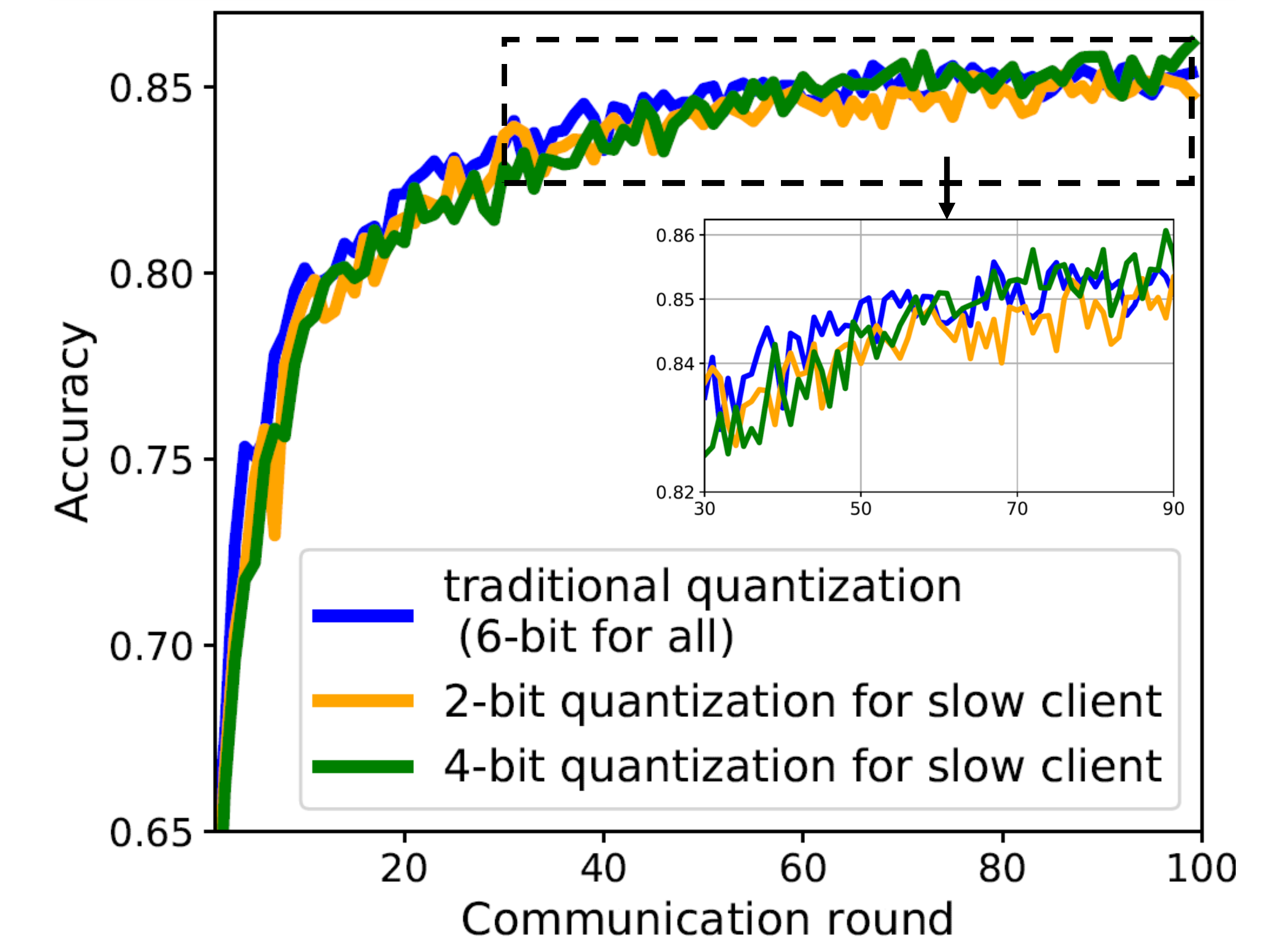}
    }
    \vspace{-0.1in}
    \caption{Different quantization strategies for heterogeneous edge devices.}
    %\vspace{-0.2in}
    \label{fig:hetero_client}
    \vspace{-0.1in}
\end{figure}

To further reduce the training time, we study the heterogeneity of the edge devices.  
Given the heterogeneous communication capability of the edge devices (clients), 
the training time depends on the slow clients with poor network conditions. 
To mitigate such straggler problem, we investigate heterogeneous gradient quantization strategies, 
which use less quantization resolutions for slower clients. 
Specifically, we train ResNet-18 on CIFAR-10 with four clients: 
three clients with data transmission rate of 20 Mbps, and one client (straggler) with data transmission rate of 5 Mbps.
We evaluate the traditional quantization strategy that uses 6-bit quantization for all clients, and compare it with four heterogeneous quantization strategies by 
letting the slowest client use 2-bit, 3-bit, 4-bit, and 5-bit quantization, respectively.

Fig.~\ref{fig:hetero_client}(a) shows the total training time of different quantization strategies, 
to reach the same accuracy of 85.0\% (near convergence). 
We observe that 3-bit, 4-bit and 5-bit quantization strategies all outperform the traditional approach, 
and the 4-bit quantization strategy has the lowest training time. 
To find out how the heterogeneous quantization strategies reduce the training time, 
we draw accuracy as a function of training rounds for different strategies, as shown in Fig.~\ref{fig:hetero_client}(b).
From the figure, we observe that 2-bit quantization takes 76 communication rounds to reach the accuracy of 85\%, while the traditional quantization takes 51 rounds. However, in the 2-bit quantization strategy, less quantization is used and each round takes less time. As a result, it has similar accumulated time to reach 85\% accuracy as that of the traditional quantization (as shown in Fig.~\ref{fig:hetero_client}(a)).
Although the 4-bit quantization strategy has 6 more rounds than the traditional quantization strategy, it can significantly reduce the training time since each round takes less time.  
In summary, we should consider both per-round communication time and the number of communication rounds 
when determining the quantization resolution for heterogeneous clients. 

These evaluation results show the potential of using adaptive and heterogeneous quantization 
to reduce the training time without compromising the model accuracy.
However, it is hard to quantify the relationship between the quantization resolution and the training time. 
For example, clients may have various transmission rates and it is hard to know which clients are the bottleneck at 
which time, and thus it is a challenge to assign quantization resolutions to heterogeneous clients to minimize 
the overall training time.
In the next section, we propose AdaGQ, an adaptive and heterogeneous gradient quantization algorithm that exploits online learning to adaptively adjust the quantization resolutions based on the gradient norms and the local training and transmission time of the clients.

%********** Method ***************

%\vspace{-0.07in}
\section{Design of AdaGQ}
\label{sec:method}
%\vspace{-0.03in}

The main challenges of designing AdaGQ are: (1) How to integrate gradient norm with the algorithm to minimize the total training time. To address this challenge, AdaGQ dynamically adjusts the number of quantization levels assigned to the clients based on the observed change of gradient norm. Specifically, when observing a larger gradient norm, AdaGQ tends to increase the number of quantization levels to preserve the precision of the gradients to reduce the number of training rounds; while for a smaller gradient norm, AdaGQ assigns fewer numbers of quantization levels to the clients to reduce the communication time, and thus reducing the total training time.
(2) How to deal with the training time bottleneck brought by the slowest clients (i.e., straggler effects). To address this challenge, AdaGQ assigns different numbers of quantization levels to different clients based on their computation and communication resources.
Intuitively, slow clients (i.e., with less resources) are assigned relatively fewer number of quantization levels to reduce the communication time to mitigate the straggler effects, while fast clients are assigned more quantization levels 
to reduce the precision loss due to quantized gradients, and then to reduce the number of training rounds. 
In the following, we first give an overview of AdaGQ, and then present the details of AdaGQ.

\subsection{Overview of AdaGQ}

\begin{figure}
    \centering
    \includegraphics[width=3.3in]{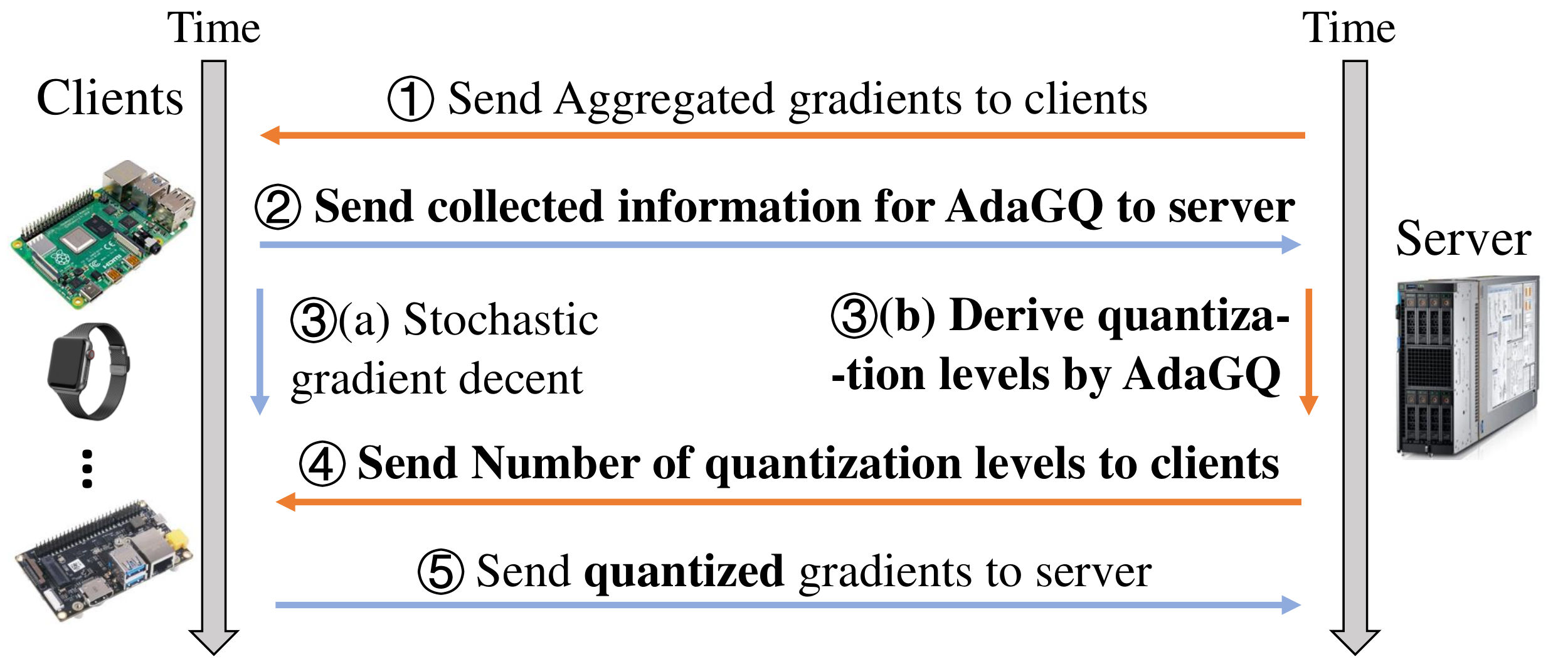}
    \vspace{-0.1in}
    \caption{Overview of AdaGQ.}
    \label{fig:system}
    \vspace{-0.1in}
\end{figure}

AdaGQ follows the system design of the state-of-the-art FL system \cite{bonawitz-mlsys2019, bonawitz-sigsac2017} 
by adopting the adaptive and heterogeneous quantization.
Fig.~\ref{fig:system} gives an overview of AdaGQ. 
In step 1, the server sends clients the aggregated gradients collected in the last round 
to synchronize the saved model parameters. In step 2, the clients collect necessary inputs of the AdaGQ algorithm, 
e.g., the losses achieved by the model when updated by gradients of different quantization levels and their corresponding training time, which are then sent to the server.
In step 3 (a), the clients apply stochastic gradient descent to the updated model (with the aggregated gradients received in step 1) and obtain the gradients. Meanwhile, in step 3 (b), the server derives the number of quantization levels for each client with the collected information in step 2. In step 4, the server sends to each client its own number of quantization levels in this round. Finally, each client quantizes the gradients derived in step 3 (a) and sends them back to the server. 

We emphasize the novel parts in the AdaGQ design with bold fonts in Fig.~\ref{fig:system}. Note that AdaGQ collects the necessary algorithm inputs in step 2 and follows an algorithm in step 3 to derive quantization levels for all the clients. More specifically, AdaGQ algorithm first determines the average quantization level of all clients in the current round based on both the loss decrease rate and the change of the gradient norm to facilitate adaptive quantization.
Then, AdaGQ derives the quantization levels for heterogeneous clients. 
In the following, we present the adaptive and heterogeneous quantization in detail, respectively.

%\vspace{-0.1in}
\subsection{Adaptive Quantization}

AdaGQ adpats the average number of quantization levels of all clients to minimize the total training time in two steps: (i) to maximize the {\em loss decrease rate}, and (ii) to calibrate the adaptation in (i) based on the change of the gradient norm.

Let $s_{i,k}$ denote the number of quantization levels used by client $i$ at round $k$, 
let $s_{k}$ denote the average number of quantization levels at round $k$,  
i.e., $s_{k} = \frac{1}{n}\sum_{i=1}^{n} s_{i,k}$, where $n$ is the number of clients.
Note that $s_{k}$ is introduced to assist the design of adaptive quantization and it does not have to be an integer.
In the first step, we adapt $s_{k}$ to minimize the training time by finding the optimal average number of quantization level $s^*_{k}$.
Then, we optimize the {\em loss decrease rate} of each round, defined as 
\begin{equation}
\label{eq:R_k}
    R_{k} = (L_{k-1} - L_k)/T_{k-1, k},
\end{equation}
where $L_k$ denotes the average loss of all clients achieved at the end of round $k$, 
and $T_{k-1, k}$ denotes the elapsed time between the end of round $(k-1)$ and that of round $k$. 
Suppose $R^*_{k}$ is the loss decrease rate achieved by $s^*_{k}$, we first construct the loss function 
\begin{equation}
    f(s_{k}) = R^{*}_{k}-R_{k}.
\end{equation}
and then explore the idea of online gradient descent based algorithms to use the derivative of the loss function to indicate the direction of the optimal solution, as shown below:
\begin{equation}
\label{eq:s_derivative}
    s_{k+1} = s_{k} - \lambda \nabla f(s_{k}),
\end{equation}
where $\lambda$ is the step size (i.e., learning rate) to update $s_{k}$, and $\nabla f(s_{k})$ denotes the derivative of the loss function at $R_{k}$.
In practice, it is impossible to obtain the exact value of the derivative $\nabla f(s_{k})$ due to the unknown form of $f(s_{k})$. 
Thus, we obtain the sign of the derivative $\nabla f(s_{k})$ which indicates the update direction, instead of the exact value.

In order to obtain the sign of $\nabla f(s_{k})$, besides the current used $s_{k}$, we use another quantization level $s^{\prime}_{k}$, which is slightly lower than $s_{k}$, and record the loss decrease rate $R^{\prime}_{k}$ achieved by $s^{\prime}_{k}$. Then, the sign of $\nabla f(s_{k})$ is derived as
\begin{equation}
\label{eq:sign}
    sign(\nabla f(s_{k})) = sign( \frac{R^{\prime}_{k}-R_{k}}{s_{k} - s^{\prime}_{k}}).
\end{equation}
The details of obtaining $R^{\prime}_{k}$ will be explained in Section~\ref{sec:adagqimp}. After obtaining the derivative sign, our algorithm updates $s_{k}$ to the opposite direction of the sign. That is
\begin{equation}
\label{eq:s_update}
    \left\{\begin{array}{ll}
         & \hat{s}_{k+1} = s_{k} - \lambda_{1}, \quad if \quad sign(\nabla f(s_{k})) = 1 \\
         & \hat{s}_{k+1} = s_{k} + \lambda_{2}, \quad if \quad sign(\nabla f(s_{k})) = -1.
    \end{array}
    \right.
\end{equation}
where $\lambda_1$ is set as half of $s_k$ so that $\hat{s}_{k+1}=s_k/2$ has one fewer bit than $s_k$, and $\lambda_2$ is set as the same of $s_k$ so that $\hat{s}_{k+1}=s_k\times 2$ has one more bit than $s_k$. 
Note that $\lambda_1$ and $\lambda_2$ are not designed as constants, and AdaGQ will explore a larger range of $s_k$, by increasing or decreasing the number of bits by 1 at a time, and quickly approach to a better setting.

We calibrate $s_{k+1}$ with the change of the gradient norm. We estimate the change of the gradient norm from round $k$ to $(k+1)$ by the observed gradient norm change from round $(k-1)$ to $k$. 
We increase $s_{k+1}$ when expecting a rise of the gradient norm and decrease $s_{k+1}$ otherwise. 
By denoting the aggregated quantized gradients by the server at the end of round $k$ as $\mathbf{g}_{k}$ and its norm as $||\mathbf{g}_{k}||$, we calibrate $\hat{s}_{k+1}$ to be $s_{k+1}$ by,
\begin{equation}
\label{eq:s_update_norm}
s_{k+1} = \hat{s}_{k+1} +\lambda_{\mathbf{g}}(\log_2||\mathbf{g}_{k}|| - \log_2||\mathbf{g}_{k-1}||)
\end{equation}
where $\lambda_{\mathbf{g}}$ is the learning rate to weight gradient norm change.

\subsection{Heterogeneous Quantization}
\label{sec:heterogeneous-quantization}

The goal of heterogeneous quantization is to let the server receive the gradients of each client 
at similar times, so that the waiting time of fast clients and server is minimized.
We first derive the relationships of the number of quantization levels among clients.

For a client $i$, its {\em local time} $t^{r}_{i,k}$ in the training round $k$ consists of its local training time $t^{cp}_{i,k}$ spent on stochastic gradient descent to derive new gradients, and the communication time $t^{cm}_{i,k}$ spent on sending the quantized gradients to the server.
% Let $s_{i,k}$ denote the number of quantization levels used by client $i$ at round $k$, 
Let $b_{i,k}$ denote the number of bits for a quantized gradient (referred to as quantization bit), which means $b_{i,k} = \lfloor \log_2(s_{i,k})+1 \rfloor$. 
The server determines the number of quantization level for each client in the round $(k+1)$ as follows. 
\begin{equation}\small
\begin{aligned}
\label{eq:localtime}
\mathbb{E}(t^{r}_{i,k+1})  = \mathbb{E}(t^{cp}_{i,k+1} + t^{cm}_{i,k+1}) 	\approx \mathbb{E}(t^{cp}_{i,k+1}) +  b_{i,k+1}\times \mathbb{E}( \frac{P}{r^{trans}_{i,k+1}}),
\end{aligned}
\end{equation}
where $P$ denotes the number of gradients which is a constant (same for all clients in all rounds), and $r^{trans}_{i,k+1}$ denotes the transmission rate of client $i$ in round $(k+1)$.
Then, our goal is to make the expected local time of each client as similar as possible,
i.e., to satisfy the condition $\mathbb{E}(t^{r}_{1,k+1}) = \mathbb{E}(t^{r}_{2,k+1}) = \cdots = \mathbb{E}(t^{r}_{n,k+1})$.
Note that we omit the time for the server to broadcast the aggregated gradients to clients since it is relatively small.
By introducing Eq.~(\ref{eq:localtime}) as the condition, for any two clients $i$ and $j$, their quantization bits $b_{i,k+1}$ and $b_{j,k+1}$ should satisfy the following:  
\begin{equation}\small
    b_{j,k+1} = \frac{1}{ \mathbb{E}(\frac{P}{r^{trans}_{j,k+1}})}(\mathbb{E}(t^{cp}_{i,k+1})-\mathbb{E}(t^{cp}_{j,k+1})+b_{i,k+1}\times \mathbb{E}(\frac{P}{r^{trans}_{i,k+1}}))
\end{equation}

\textbf{Local training time and transmission rate estimation.} In practice, in order to assign $b_{i,k+1}$ to every client, we have to estimate the local training time $\mathbb{E}(t^{cp}_{i,k+1})$ and the transmission time coefficient $\mathbb{E}(\frac{P}{r^{trans}_{i,k+1}})$.
Since the per-round local training time of a client does not vary much, $\mathbb{E}(t^{cp}_{i,k+1})$ is estimated by the average of all the historical local training times spent by client $i$, i.e., $\mathbb{E}(t^{cp}_{i,k+1}) = \frac{1}{k} \sum_{k^\prime=1}^{k} t^{cp}_{i,k^\prime}$.

On the other hand, the transmission rate may have variations over different training rounds (but usually smooth) and thus we estimate $\mathbb{E}(\frac{P}{r^{trans}_{i,k+1}})$ based on the same transmission rate of last round, i.e., $\mathbb{E}(\frac{P}{r^{trans}_{i,k+1}}) \approx \frac{P}{r^{trans}_{i,k}}=t^{cm}_{i,k}/b_{i,k}$.
Given the number of quantization bits (levels) of one client (e.g., client $i$), the number of quantization bits of other clients can be determined as follows. 
\begin{equation}\small
\label{eq:bi}
\begin{aligned}
\small
b_{j,k+1} = \frac{ b_{j,k}}{t^{cm}_{j,k}} (\frac{1}{k} \sum_{k^\prime=1}^{k} t^{cp}_{i,k^\prime} - \frac{1}{k} \sum_{k^\prime=1}^{k-1} t^{cp}_{j,k^\prime} + b_{i,k+1}\times \frac{t^{cm}_{i,k}}{ b_{i,k}}), \\
\forall j \in \{1,\cdots,n\}, j \neq i.
\end{aligned}
\end{equation}
where $b_{i,k+1} = \lfloor \log_2(s_{i,k+1})+1 \rfloor$, for $i=1,2,...,n$, and $\frac{1}{n}\sum^n_i{s_{i,k+1}}=s_{k+1}$. 
Thus, we can derive $b_{i,k+1}$ from Eq. (\ref{eq:bi}) and refine $s_{i,k+1}$ as $(2^{b_{i,k+1}}-1)$. Once the server 
determines the number of quantization levels for client $i$ (i.e., $s_{i,k+1}$), it sends $s_{i,k+1}$ to client $i$ as its quantization in round $(k+1)$.
%Since in practice $b_{i,k+1}$ should be an integer, the client $i$ takes the floor of the  $b_{i,k+1}$, if the result is not an integer, for the quantization.

\subsection{Implementation of AdaGQ}
\label{sec:adagqimp}

% figure timeline
\begin{figure}
    \centering
    \includegraphics[width=3.35in]{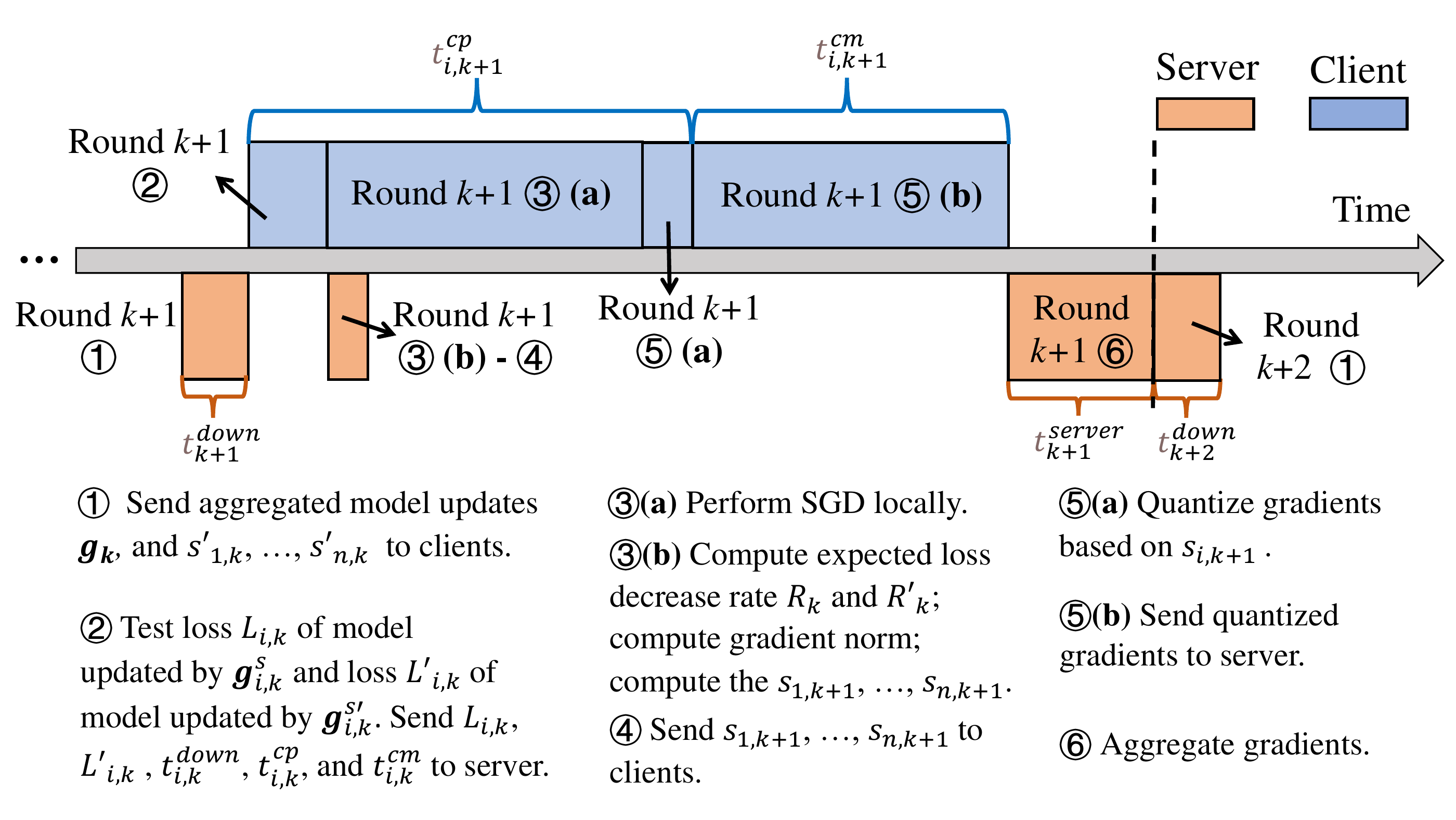}
    \vspace{-0.1in}
    \caption{Timeline in Round $(k+1)$ of AdaGQ.}
    \label{fig:adaptive_timeline}
    \vspace{-0.1in}
\end{figure}

As an example, we describe how our algorithm runs in a round $(k+1)$. 
As shown in Fig.~\ref{fig:adaptive_timeline}, at the beginning of round $(k+1)$, the server broadcasts the aggregated gradients $\mathbf{g}_{k}$ obtained in the last round ($k$), and a parameter $s^\prime_{i,k}$ (introduced later) to the clients (step 1).  
Next, to estimate $sign(\nabla f(s_{k}))$, AdaGQ has to estimate $R_{k}$ and $R^{\prime}_{k}$ based on Eq.(\ref{eq:sign}). Recall that the loss decrease rate $R_{k}$ is defined in Eq.~(\ref{eq:R_k}). Thus, AdaGQ needs to estimate the average loss $L_{k-1}$, $L_k$ and the round time $T_{k-1, k}$ to derive $R_{k}$. In addition, AdaGQ needs to estimate $R^{\prime}_{k}$, which is the loss decrease rate if $s^\prime_{k}$ was used instead of $s_k$ in the round $k$, requiring 
further estimation of $L^\prime_{k}$ and $T^\prime_{k-1, k}$. 
Because these average losses and the round times may not be easily measured, AdaGQ estimates their values in step 2.
After client $i$ receives the aggregated gradients $\mathbf{g}_k$ from the server, it quantizes $\mathbf{g}_k$ with $s_{i,k}$ (i.e., the number of quantization level assigned to client $i$ with $s_k$) and $s^{\prime}_{i,k}$ (i.e.,the number of quantization level assigned to client $i$ with $s^\prime_k$) quantization levels, respectively. 
Here $s^\prime_{k}$ is an auxiliary selected by the server by $s^\prime_{k} = \lfloor s_{k}/2\rfloor$ (i.e., one bit fewer than $s_k$), and $s^\prime_{i,k}$ is derived from $s^\prime_k$ following the same way in which the server derived $s_{i,k}$ from $s_k$ in round $k$.
Suppose the obtained quantized gradients are denoted as $\mathbf{g}^{s}_{i,k}$ and $\mathbf{g}^{s^\prime}_{i,k}$. 
The client $i$ computes two losses $L_{i,k}$ and $L^{\prime}_{i,k}$, which are the losses obtained by the models updated with $\mathbf{g}^{s}_{i,k}$ and $\mathbf{g}^{s^\prime}_{i,k}$, respectively.
% Note that this step does not incur additional data transmission. The client computes $\mathbf{g}^{s}_{i,k}$ and $\mathbf{g}^{s^\prime}_{i,k}$ locally based on the received gradients. 
Next, client $i$ uploads the calculated losses $L_{i,k}$ and $L^{\prime}_{i,k}$, 
as well as the parameters downloading time $t_{i,k}^{down}$, local computation time $t^{cp}_{i,k}$ and communication time $t^{cm}_{i,k}$ in round $k$, to the server. Then, the clients conduct a new round of model training SGD (step 3 (a)).

After the server receives the information from all clients, at the same time with step 3 (a), 
it needs to compute two estimated loss decrease rates $R_k$ and $R^{\prime}_k$ (step 3 (b)). 
The server first derives the estimation of the average loss $L_k$ by averaging the loss $L_{i,k}$ collected from all the clients, i.e., $\bar{L}_{k} = \frac{1}{n}\sum_{i=1}^{n} L_{i,k}$ where $n$ is the number of the clients.
Similarly, the server estimates $L^\prime_k$ by $\bar{L}^{\prime}_{k} = \frac{1}{n}\sum_{i=1}^{n} L^{\prime}_{i,k}$. As the server saves the status of model parameters and gradients at the beginning of round $k$, it can easily obtain the real loss $L_{k-1}$ for Eq.~(\ref{eq:R_k}).

The server estimates the average time of training round $k$ when clients quantize the gradients under the condition of $s_{k}$ and $s^{\prime}_{k}$, denoted as $T_{k-1,k}$ and $T^{\prime}_{k-1,k}$, respectively.
$T_{k-1,k}$ is determined by the slowest client in round $k$, i.e., the maximum time spent among all clients, and it is  obtained by the server as follows:
\begin{equation}
    T_{k-1,k} = \mathop{max}\limits_{i}\{t^{cp}_{i,k} + t^{cm}_{i,k} + t^{down}_{i,k}\} + t^{server}_{k}.
\end{equation}
To estimate $T^{\prime}_{k-1,k}$, the main challenge is to estimate the transmission time ${t'}^{cm}_{i,k}$ since the main change is the number of bits for transmission. To bridge this gap, AdaGQ computes the number of bits when using $s_{i,k}$ by $b_{i,k} = \lfloor \log_2(s_{i,k})\rfloor + 1 $, and that of using $s^\prime_{i,k}$ by $b^\prime_{i,k} = \lfloor \log_2(s^\prime_{i,k})\rfloor + 1 $. Then the transmission time ${t'}^{cm}_{i,k}$ can be estimated as $\frac{b^\prime_{i,k}}{b_{i,k}}{t}^{cm}_{i,k}$, and thus AdaGQ estimates the training round time $T^{\prime}_{k-1,k}$ of using $s^{\prime}_{k-1}$ as follows.
\begin{equation}
\small
    T^{\prime}_{k-1,k} = \mathop{max}\limits_{i}\{t^{cp}_{i,k} + \frac{\lfloor \log_2(s^{\prime}_{i,k})\rfloor+1}{\lfloor \log_2(s_{i,k})\rfloor+1} t^{cm}_{i,k}+ t^{down}_{i,k}\}+ t^{server}_{k}.
\end{equation}
Then, the server estimates the two loss decrease rates
$R_{k}$ and $R^{\prime}_{k}$ with Eq.~(\ref{eq:R_k}) as follows.
\begin{equation}
\label{eq:R_k_imp}
    R_{k} = (L_{k-1} - \bar{L}_{k})/T_{k-1, k}, \quad R^{\prime}_{k} = (L_{k-1} - \bar{L}^{\prime}_{k})/T^{\prime}_{k-1, k}
\end{equation}
The server estimates the sign of $\nabla f(s_{k-1})$ following Eq.~(\ref{eq:sign}) using $R_k$ and $R^\prime_{k}$.
To obtain $s_{k+1}$, the next step in step 3 (b) is to compute the gradient norm of the aggregated gradients (i.e., $\mathbf{g}_k$) and update $s_{k+1}$ following Eq.~(\ref{eq:s_update}) and Eq.~(\ref{eq:s_update_norm}).
Finally, the server derives $b_{i,k+1}$ and $s_{i,k+1}$, for $i=1,2,\cdots,n$, as introduced in Section 
\ref{sec:heterogeneous-quantization}.

After the client $i$ receives $s_{i,k+1}$ from the server (step 4) and finishes computing the new model gradients in the current round, client $i$ quantizes the newly computed gradients in $s_{i,k+1}$ quantization levels. The quantized gradients are then uploaded to the server (step 5). Finally, the server collects all quantized gradients from all clients, 
and conducts a global aggregation on these gradients to generate $\mathbf{g}_{k+1}$ in step 6. 
The server also prepares $s^\prime_{k+1}$ as $\lfloor s_{k+1}/2\rfloor$, 
derives $s^\prime_{i,k+1}$ for $i=1,2,...,n$, and sends it to the clients in the next round, i.e., round $(k+2)$.
The details are shown in Algorithm 2.

% ***** AdaGQ ******

\setlength{\textfloatsep}{0.02cm}
\begin{algorithm}[t]
\SetAlgoLined
\textbf{Initialization:} global model weight $\mathbf{w}_0$; initial model weights of clients $\mathbf{w}_{i,0} = \mathbf{w}_0, \forall i \in \{1, \cdots, n\}$; initial number of quantization levels $s_0$; $s_{i,0} = s_{0}, \forall i \in \{1, \cdots, n\}$.\\

% $\mathbf{w}_k \gets$ Download the latest global model from server;\\
\For{ each $k=1,2,\cdots$}
 {
    \For{each client $i = 1, 2, \cdots, n$ in parallel}
    {
        $\mathbf{g}_{k} \gets$ receives aggregated gradients from server;\\
        $\mathbf{w}_{i,k+1} \gets$ update model parameters with $\mathbf{g}_{k}$;\\
        $\mathbf{g}^s_{i,k}$, $\mathbf{g}^{s^\prime}_{i,k}$ $\gets$ quantize gradients;\\
        $\mathbf{w}^\prime_{i}$, $\mathbf{w}^{\prime\prime}_{i}$ $\gets$ model parameters when updated with $\mathbf{g}^s_{i,k}$, $\mathbf{g}^{s^\prime}_{i,k}$;\\
        $L_{i,k}$, $L^{\prime}_{i,k}$ $\gets$ losses of $\mathbf{w}^\prime_{i}$ and $\mathbf{w}^{\prime\prime}_{i}$ in local test set;\\
        Send $L_{i,k}$, $L^{\prime}_{i,k}$, $t^{down}_{i,k}$, $t^{cp}_{i,k}$, $t^{cm}_{i,k}$ to server;\\
        $g(\mathbf{w}_{i,k+1}) \gets$ Perform SGD locally;\\
        Receive $s_{i,k+1}$ from server;\\
        $Q_{s_{i,k+1}}(\mathbf{g}_{i,k+1})$ $\gets$ Perform gradient quantization;\\
        Upload $Q_{s_{i,k+1}}(\mathbf{g}_{i,k+1})$ to server;\\
        record $t^{down}_{i,k+1}$, $t^{cp}_{i,k+1}$, $t^{cm}_{i,k+1}$ of this round;\\
    }
    
    \textbf{The server does:}\\
    {
     \quad Send the aggregated gradients $\mathbf{g}_k$ together with $s^{\prime}_{i,k}$, $i=1,2, \cdots,n$ to the clients;\\
     \quad Receive $L_{i,k}$, $L^{\prime}_{i,k}$, $t^{down}_{i,k+1}$, $t^{cp}_{i,k}$, $t^{cm}_{i,k}$ from all clients;\\
     \quad Compute the expected loss decrease rate $R_{k}$, $R^{\prime}_{k}$ following Eq.~(\ref{eq:R_k_imp});\\
     \quad Compute $s_{k+1}$ following Eq.~(\ref{eq:s_update_norm});\\
     \quad Compute $s_{i,k+1}$ following Eq.~(\ref{eq:bi}) and send $s_{i,k+1}, i = 1,2,\cdots,n$ to clients;\\
     
     \quad Gather $Q_{s_{i,k+1}}(\mathbf{g}_{i,k+1}), i=1,2, \cdots,n$;\\
     \quad $||\mathbf{g_{k+1}}||$ $\gets$ Computes gradient norm;\\

     \quad  $\mathbf{g}_{k+1}$ $\gets$ Aggregate gradients by $\sum_{i=1}^n p_i Q_{s_{i,k+1}}(\mathbf{g}_{i,k+1})$;\\
     \quad  Set $s^\prime_{k+1}$ as $\lfloor s_{k+1}/2\rfloor$ and derive $s^\prime_{i,k+1}$ for $i=1,2,...,n$ based on Eq.~(\ref{eq:bi}).\\
    }
}
\caption{The AdaGQ Algorithm}
\label{alg:AdaGQ}
\end{algorithm}
\setlength{\floatsep}{0.02cm}

\section{Performance Evaluations}
\label{sec:evaluation}

In this section, we evaluate AdaGQ against four baselines under four federated learning (FL) tasks. We first introduce the 
evaluation setup and then present the evaluation results of all algorithms based on four FL tasks.

\subsection{Evaluation Setup}
\label{sec:setup}

% We implemente AdaGQ with PyTorch. With the Python threading library, we simulate a large number of devices with lightweight threads, each running real-world PyTorch models.
We evaluate the proposed algorithm with non-iid data distribution at 20 clients on various learning tasks and compare its performance with state-of-the-art algorithms.

\textbf{Models and datasets.}
We consider two model architectures with different parameter sizes: ResNet-18~\cite{he2016deep} and GoogLeNet~\cite{szegedy2015going}.
ResNet-18 is a CNN network consisting of residual blocks with over 11 million parameters.
GoogLeNet is a 22-layer CNN network without any skip connections, which has over 6 million parameters.
We evaluate AdaGQ by training ResNet-18 and GoogLeNet on two benchmark datasets: Cifar-10~\cite{krizhevsky2009learning} and FashionMNIST (FMNIST)~\cite{xiao2017fashion}.
The CIFAR-10 dataset consists of 50K color images as the training set and 10K color images for testing, where each image belongs to one of the 10 classes.
The FMNIST dataset contains 60K train and 10K test grey scale images of 10 different fashion items. 
Four FL tasks are used in the evaluation, i.e., training ResNet-18 on Cifar-10, training ResNet-18 on FMNIST, training GoogLeNet on Cifar, and training GoogLeNet on FMNIST (all with cross-entropy loss function).

\begin{figure}[t!]
    % (a)
    \subfigure[ResNet-18 on Cifar-10]{
    \centering
        \includegraphics[width=1.64in]{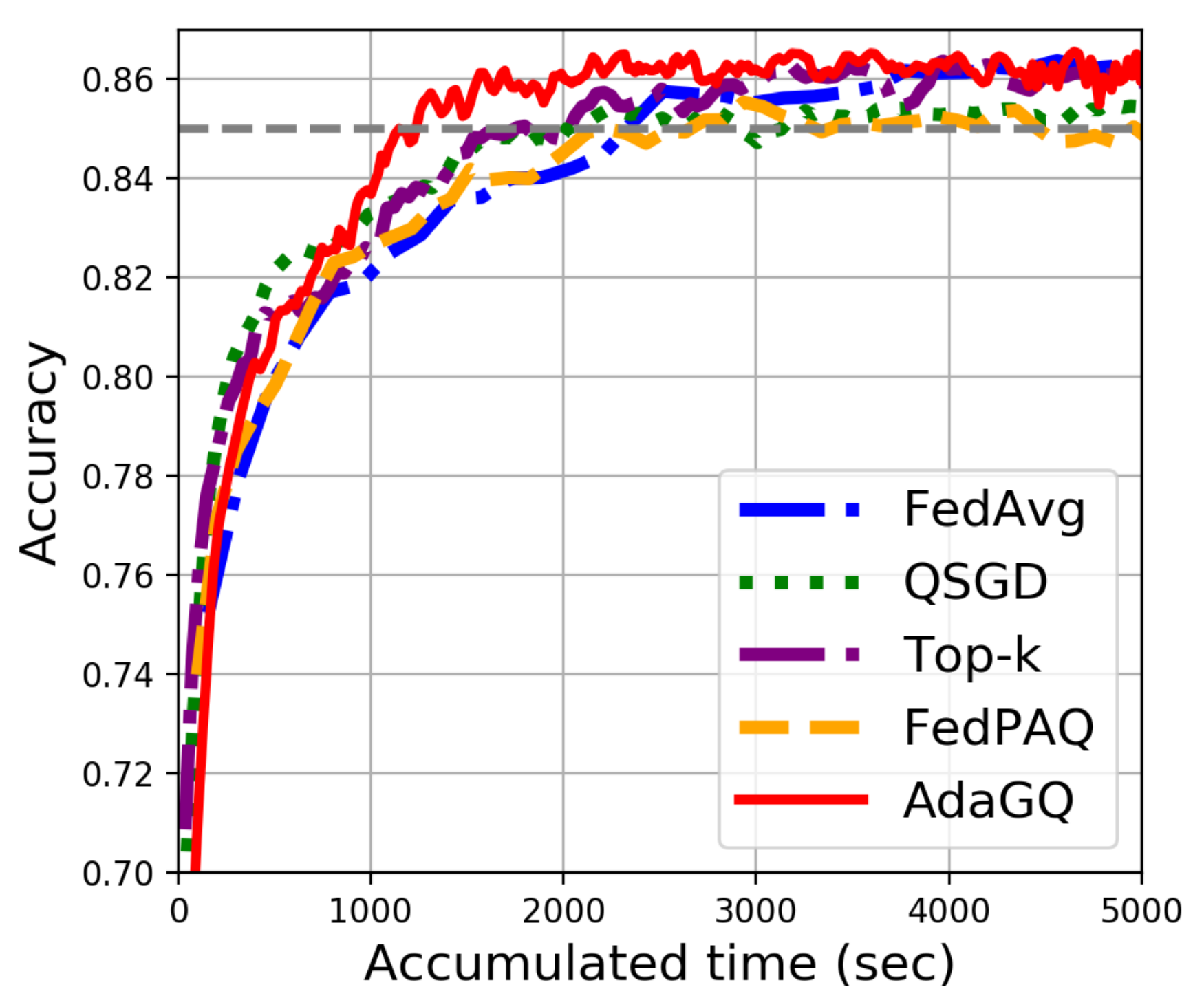}
    }
    \hspace*{-1.0em}
    % (b)
    \subfigure[ResNet-18 on FMNIST]{
    \centering
        \includegraphics[width=1.64in]{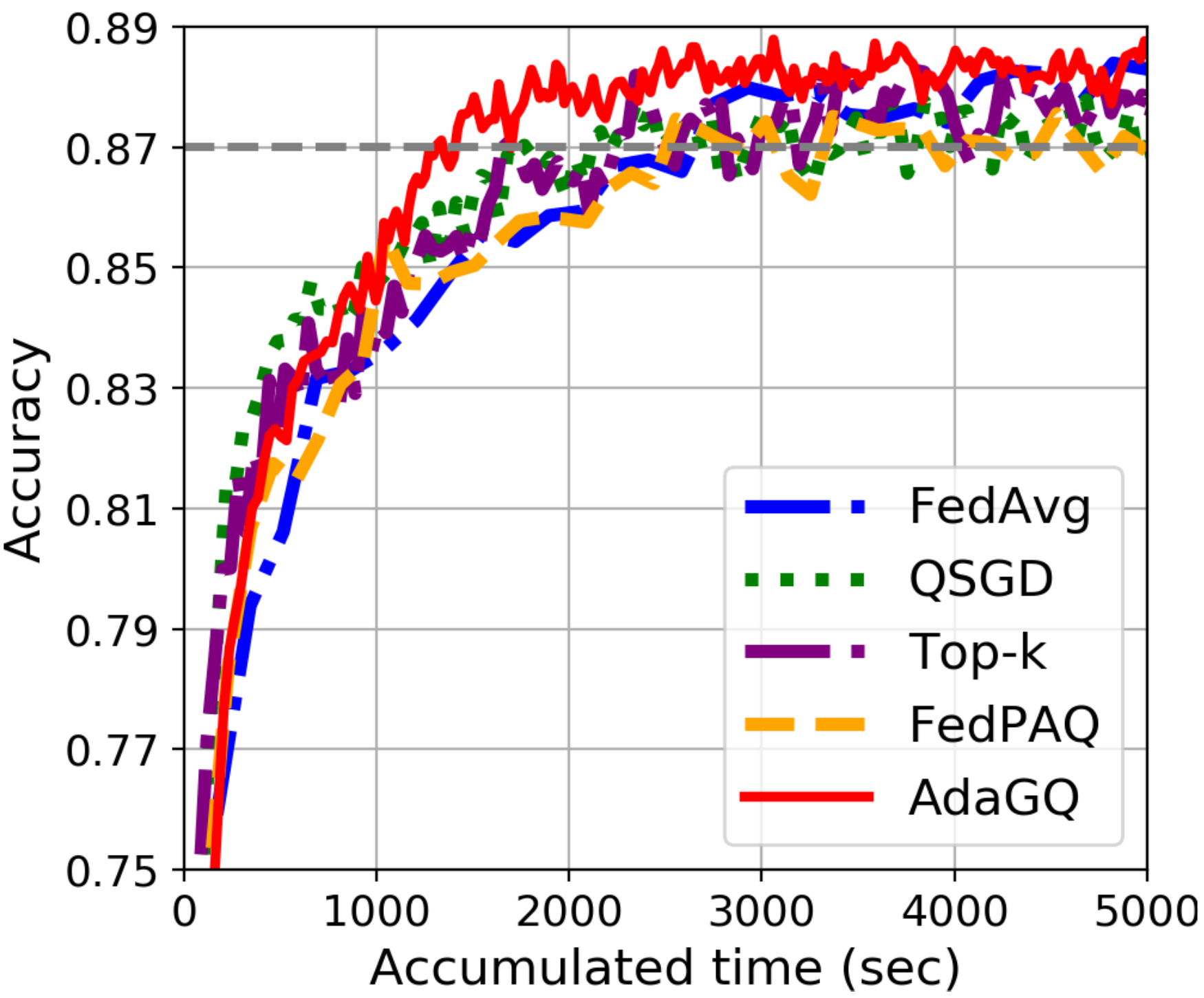}
    }
    \hspace*{-1.0em}\\
    % change line
    % (c)
    \subfigure[GoogLeNet on Cifar-10]{
    \centering
        \includegraphics[width=1.64in]{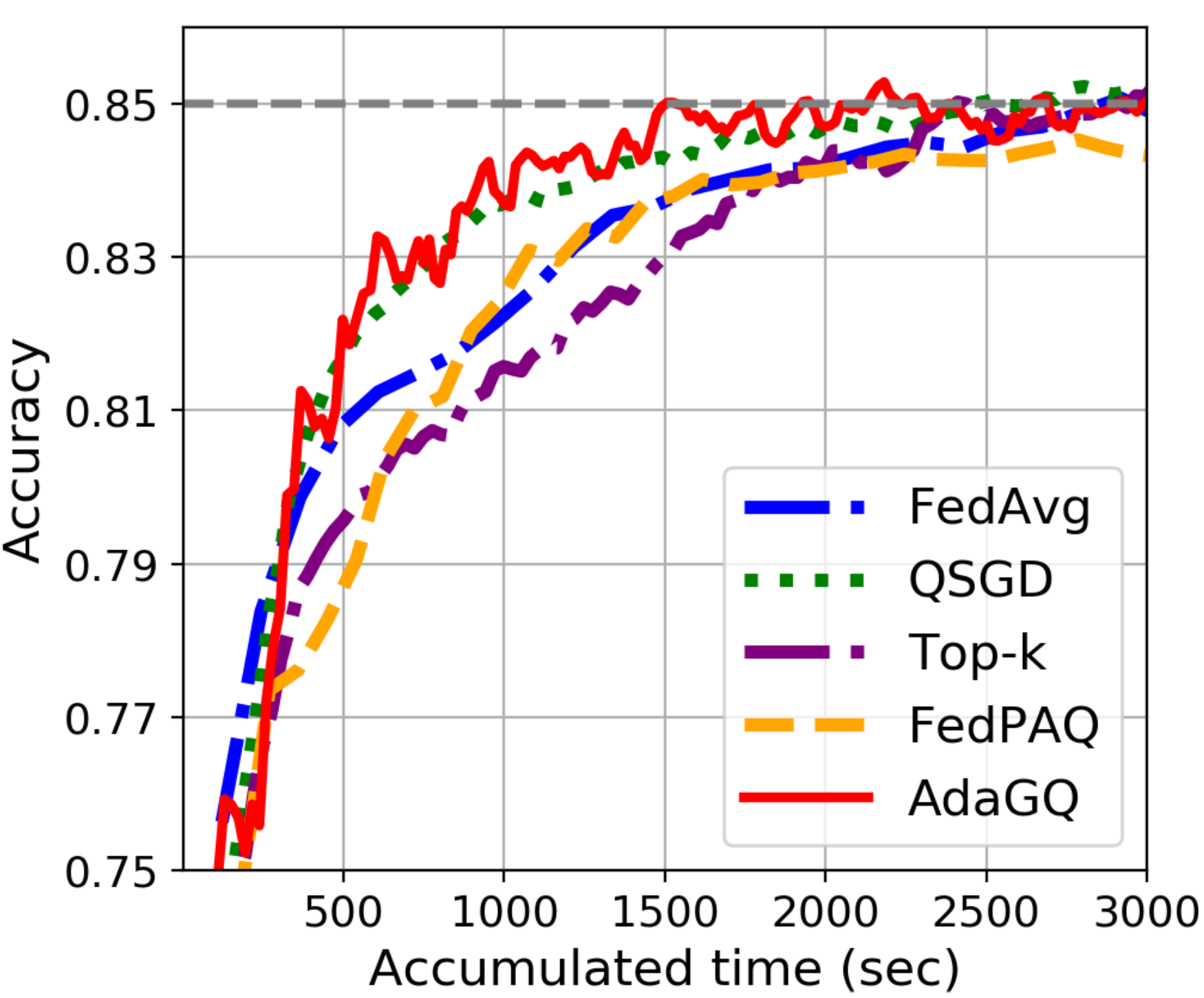}
    }
    \hspace*{-1.0em}
    % (d)
    \subfigure[GoogLeNet on FMNIST]{
    \centering
        \includegraphics[width=1.64in]{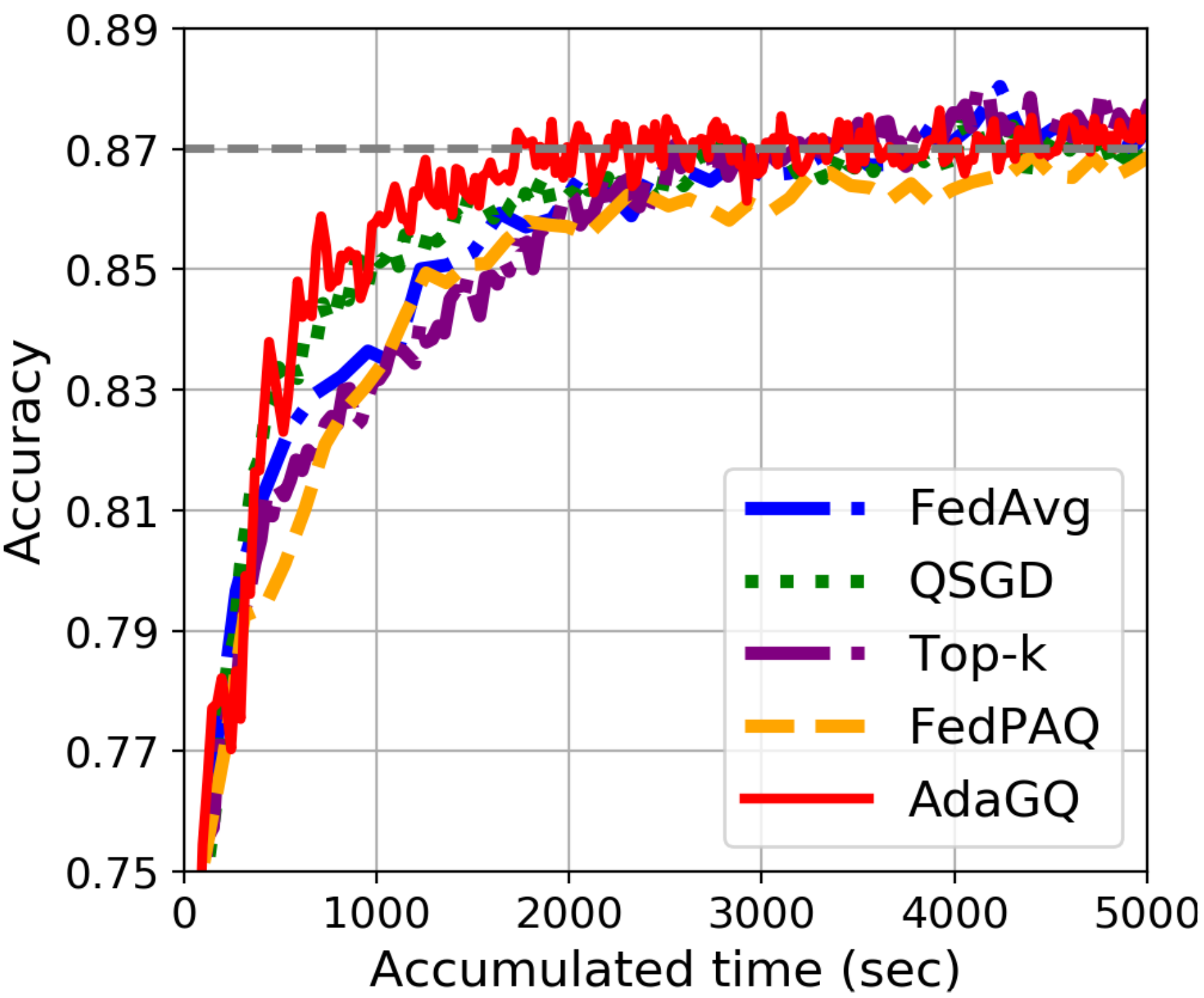}
    }
    \hspace*{-1.0em}
    % \centering
    \caption{Accuracy v.s. accumulated time of AdaGQ compared to baselines.}
    \label{fig:acc_time_compare}
\end{figure}

\begin{figure}[t!]
\vspace{-0.1in}
    % (a)
    \subfigure[ResNet-18 on Cifar-10]{
    \centering
        \includegraphics[width=1.64in]{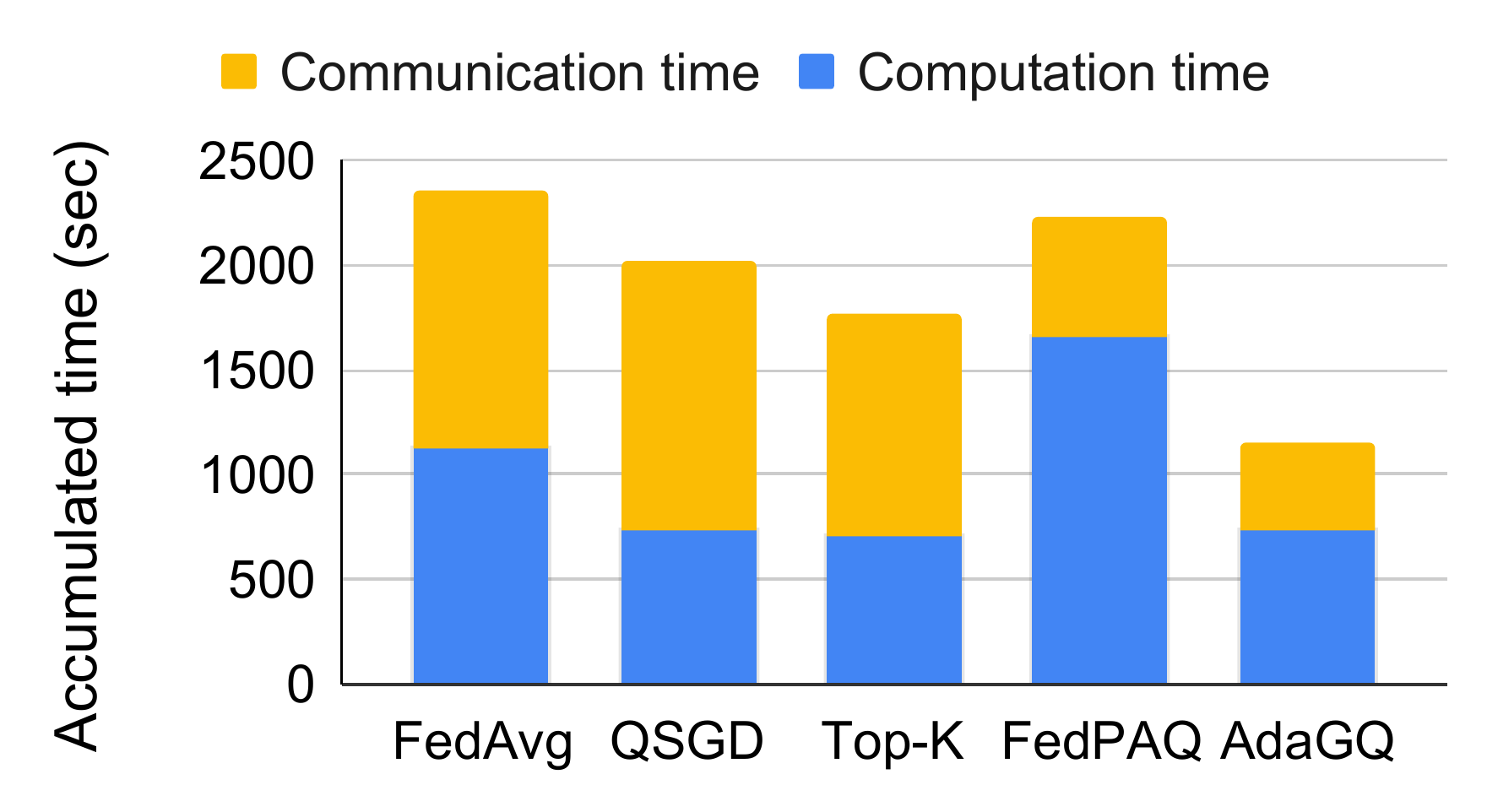}
    }
    \hspace*{-1.0em}
    % (b)
    \subfigure[ResNet-18 on FMNIST]{
    \centering
        \includegraphics[width=1.64in]{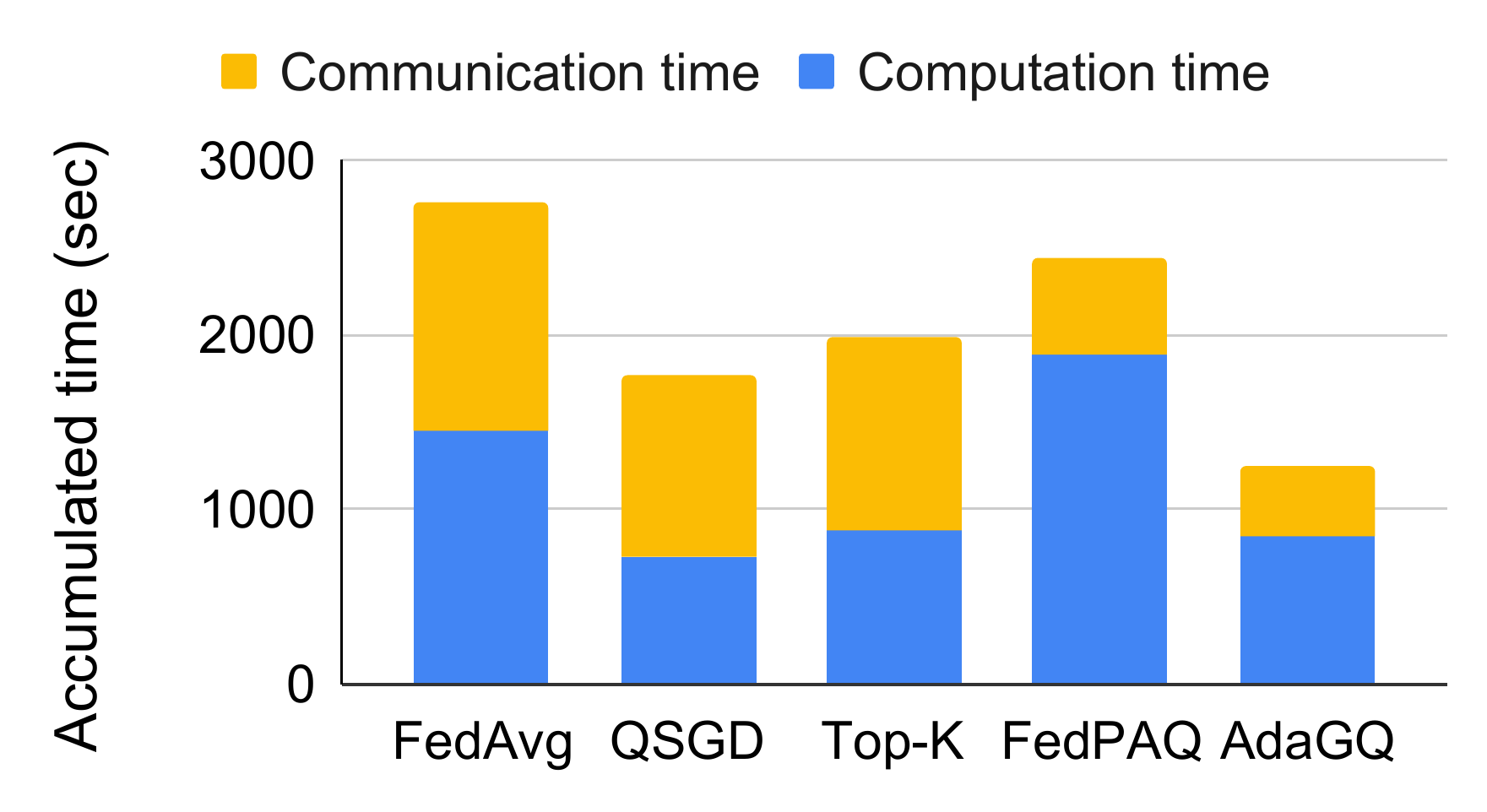}
    }
    \hspace*{-1.0em}
    
    % change line
    % (c)
    \subfigure[GoogLeNet on Cifar-10]{
    \centering
        \includegraphics[width=1.64in]{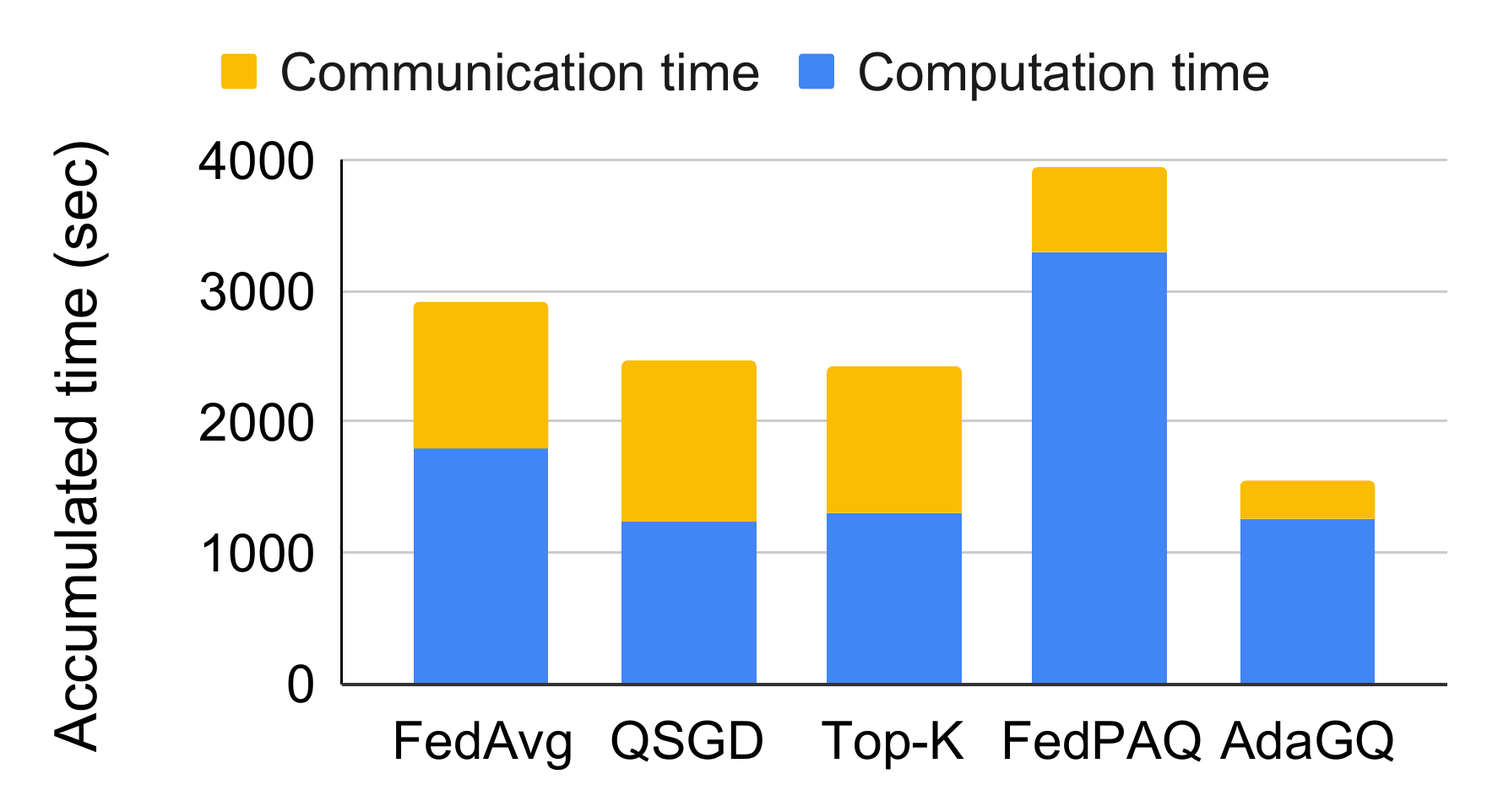}
    }
    \hspace*{-1.0em}
    % (d)
    \subfigure[GoogLeNet on FMNIST]{
    \centering
        \includegraphics[width=1.64in]{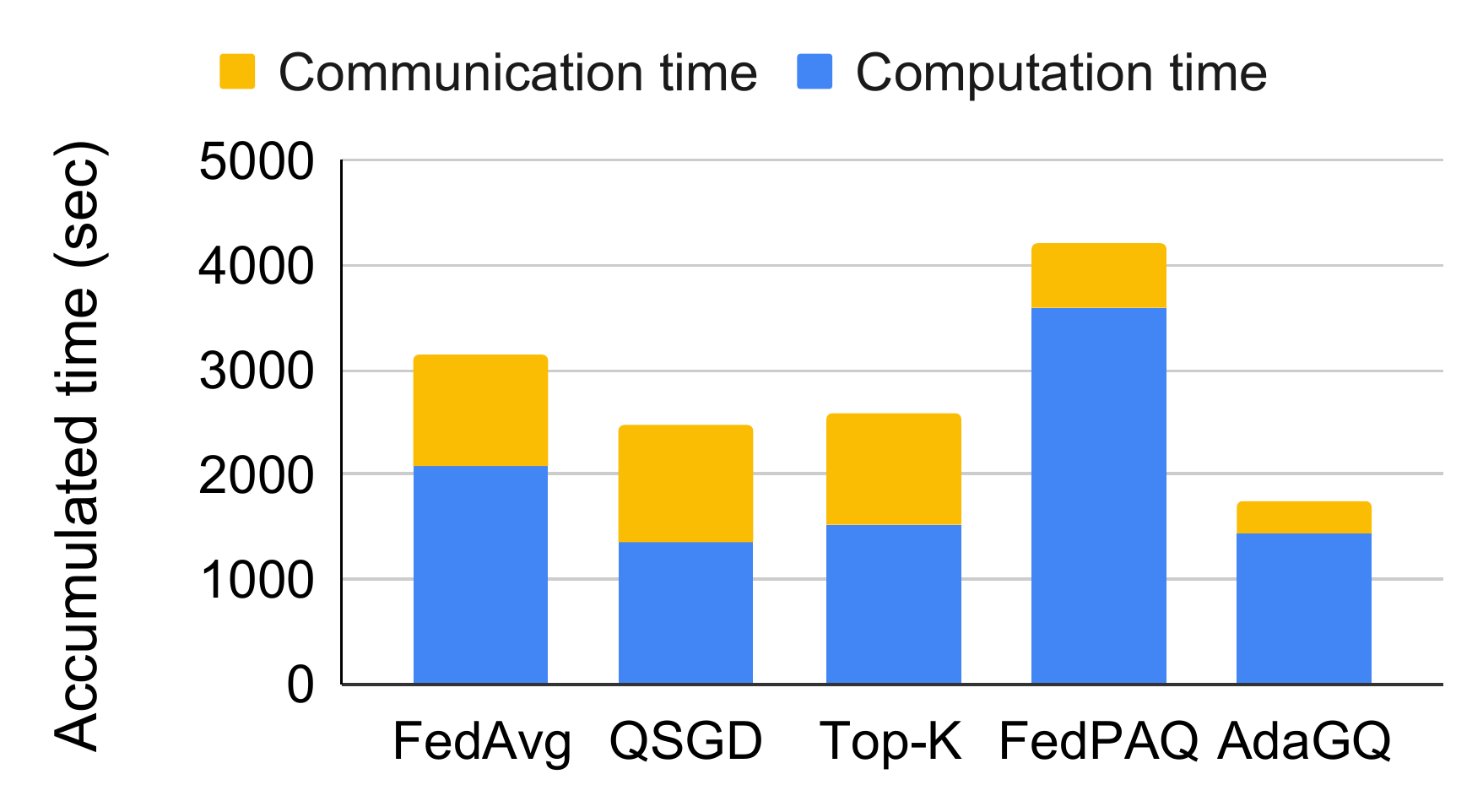}
    }
    \hspace*{-1.0em}
    \vspace{-0.1in}
    % \centering
    \caption{Total training time of AdaGQ compared to baselines.}
    \label{fig:time_compare}
\end{figure}

\textbf{Methods for comparison.}
We compare AdaGQ with the following four baseline approaches.

\begin{itemize}
    \item \textbf{FedAvg} \cite{haddadpour-nips2019}: clients communicate updated local parameters with the central server after multiple epochs of local training and download the aggregated global model. Here the communication period is set to be 5 epochs.
    
    \item \textbf{QSGD} \cite{alistarh-nips2017}: clients send quantized gradients to the central server and download the aggregated global model for every epoch. The number of quantization levels is set to be 8-bit.
    
    \item \textbf{Top-k}~\cite{aji-emnlp2017} is a sparsification method that compresses the communicated gradients by selecting the largest $k$ elements of the gradients. In this method, clients send sparse gradients to the central server and download the aggregated global model every epoch. We set $k$ to be 10\% of the total parameters.
    
    \item \textbf{FedPAQ}~\cite{reisizadeh-aistats2020} incorporates periodic averaging into QSGD. In FedPAQ, models are trained multiple epochs at clients and only periodically averaged at the server. Clients quantize their updates before uploading. 
    Similar to FedAvg, we set the communication period to be 5 epochs and the number of quantization levels to be 8-bit.
\end{itemize}

\textbf{Hyperparameters.}
As the default configuration, we set the local batch size to 32 and assign every client an equally sized subset of the training data. For each client, the data transmission rate is initialized to be a rate sampled randomly between 5 Mbps and 20 Mbps by default.
% For the data IID-ness
We set the initial learning rates for both ResNet-18 and GoogLeNet to be 0.01 and the decay as 0.995. 
% The threshold $\eta$ to detect the critical phase is set to be 0.5 as suggested in \cite{agarwal-mlsys2021}.
For AdaGQ, the initial number of quantization levels is set to be 8-bit, which is relatively large as suggested in \cite{alistarh-nips2017}.
The step size $\lambda_g$ is set to 1.
% The minimum and maximum quantization level are set to be 4 (2-bit) and 256 (8-bit) respectively.
%We set the number of groups to be four.
% data heterogeneity
% For the IID case, the data are uniformly split to every client, and every client holds all the classes. 
%For the non-IID case, every client holds data from only 50\% of the total classes.
% resource heterogeneity
%The uplink data transmission rate of each client is randomly generate from 5 Mbps to 20 Mbps. 
%The downlink data rate is set to be 20 Mbps.
% The local computation time of each client is set to be XXX % find some data in related work 

Similar to the definition in \cite{wang-infocom2020},
we use $\sigma_d$ to denote the level of non-iid data, which corresponds to the fraction of data that only belongs to one class at each client. 
For example, $\sigma_d = 0.2$ means that 20\% of the data on one client belongs to one class and the remaining 80\% of the data uniformly belongs to other classes.
For the baselines in comparison, we set their hyper-parameters (as shown above) the same as those suggested in the corresponding literature.

We evaluate all the algorithms in terms of the total \textit{wall-clock training time}, including computation time, communication time and 
all extra overhead, when they reach the target accuracy.

%\vspace{-0.07in}
\subsection{Comparison of the Algorithms}
\label{sec:comp_baseline}
%\vspace{-0.03in}
% Table 1:
% Cifar-10: The performance when reach the same accuracy
% network, noniid-level, accuracy, data sent, wall-clock time
\begin{table}[t]
    \centering
    \resizebox{\columnwidth}{!}{
    \begin{tabular}{m{0.3cm}<{\centering} m{0.9cm}<{\centering} m{0.9cm}<{\centering} m{1.5cm}<{\centering} m{3.0cm}<{\centering} }
    \hline
    \hline
    $\sigma_d$ & Method  & Avg. rounds &  Avg. data \quad uploaded & Total time (Second)\\
    \hline
    \hline
    \multirow{6}{*}{0.2}
    & FedAvg  & 13.25 & 6.52  (1$\times$) & 2190.89$\pm$ 40.43  (1$\times$) \\ \cline{2-5}
    & QSGD  & 44.50 & 5.28 (1.23$\times$)  & 1891.91$\pm$ 58.21 (1.16$\times$) \\ \cline{2-5}
    & Top-k  & 40.75 & 3.92 (1.67$\times$)  & 1568.57$\pm$ 62.43 (1.40$\times$)\\ \cline{2-5}
    & FedPAQ  & 19.25 & 2.33 (2.80$\times$) & 2014.05$\pm$ 46.32 (1.09$\times$) \\ \cline{2-5}
    % & X  & Y & X  & Y \\ \cline{2-5}
    & \textbf{AdaGQ}  & \textbf{43.25} & \textbf{3.43 (1.90$\times$)}  & \textbf{1033.50$\pm$ 49.86 (2.12$\times$)} \\ 
    \hline
    \hline
    \multirow{6}{*}{0.5}
    & FedAvg  & 14.25 & 7.01 (1$\times$)  & 2359.41$\pm$ 44.10  (1$\times$) \\ \cline{2-5}
    & QSGD  & 48.00 & 5.72 (1.23$\times$) & 2018.04 $\pm$ 61.20 (1.17$\times$)\\ \cline{2-5}
    & Top-k  & 45.75 & 4.40 (1.59$\times$) & 1759.86 $\pm$ 64.49 (1.34$\times$)\\ \cline{2-5}
    & FedPAQ  & 21.25 & 2.57 (2.73$\times$) & 2226.06$\pm$ 47.24 (1.06$\times$) \\ \cline{2-5}
    % & X  & Y & X  & Y \\ \cline{2-5}
    & \textbf{AdaGQ}  & \textbf{47.00} & \textbf{3.73 (1.88$\times$)} & \textbf{1129.60 $\pm$ 53.56 (2.09$\times$)} \\
        \hline
        \hline
    \multirow{6}{*}{0.8}
    & FedAvg  & 20.25 & 9.96 (1$\times$)  & 3370.59 $\pm$ 49.25 (1.26$\times$)\\ \cline{2-5}
    & QSGD  & 69.75 & 8.28 (1.20$\times$) & 2942.98 $\pm$ 65.32 (1.44$\times$)\\ \cline{2-5}
    & Top-k  & 60.50 & 5.82 (1.71$\times$) & 2295.50 $\pm$ 71.10 (1.85$\times$)\\ \cline{2-5}
    & FedPAQ  & 39.75 & 4.81 (2.07$\times$) & 4240.11 $\pm$ 48.20 (1$\times$)\\ \cline{2-5}
    % & X  & Y & X  & Y \\ \cline{2-5}
    & \textbf{AdaGQ}  & \textbf{68.00} & \textbf{5.40 (1.85$\times$)}  & \textbf{1634.31 $\pm$ 56.71 (2.59$\times$)} \\
    \hline
    \hline
    \end{tabular}
    }
    \caption{ResNet-18 on Cifar-10 under different $\sigma_d$}
    \label{tab:non-iid_level_resnet}
    % \vspace{-0.15in}
\end{table}

First, we compare the wall-clock time of AdaGQ with all baseline algorithms when they reach the same accuracy (with $\sigma_d=0.5$). Fig.~\ref{fig:acc_time_compare} shows the accuracy over accumulated time of the four FL tasks respectively. 
We observe that AdaGQ takes the least amount of time to reach the accuracy of 85.0\% for Cifar-10 and 87.0\% for FMNIST, reducing the training time by 29.1\%-34.8\% compared to the best baselines (i.e., Top-k in Fig.~\ref{fig:acc_time_compare}(a)(c) or QSGD in Fig.~\ref{fig:acc_time_compare}(b)(d)), and 45.5\%-52.1\% compared to FedAvg under the four FL tasks. 

Among the baselines, FedPAQ spends longer time than most of other baselines and fails to reach the target accuracy when training GoogLeNet on Cifar-10 and FMNIST. This is because FedPAQ incorporates both periodic averaging and gradient quantization, which incur more information loss in each round to delay the convergence.
%, and thus take more time to reach the same accuracy.
In addition, we observe that Top-k and QSGD spend less training time than FedAvg, which suggests that the gradient 
compression can save more time than periodic averaging.
By comparing AdaGQ with Top-k and QSGD, we observe that AdaGQ outperforms Top-k and QSGD consistently on all four FL tasks, which validates the advantages of adaptive and heterogeneous quantization.

To analyze how AdaGQ reduces the total training time, we separate the communication time and the computation time for all the algorithms, as shown in Fig.~\ref{fig:time_compare}.
We observe that AdaGQ spends similar computation time but significantly less communication time compared to QSGD (the second best algorithm). Because the computation time spent in each round is similar for both algorithms, 
having similar computation time indicates that both algorithms take similar numbers of rounds to reach the target accuracy. 
However, AdaGQ saves the communication time in each round, by adjusting the number of quantization levels based on the adaptive and heterogeneous quantization, and thus reduces the accumulated wall-clock time.
In addition, among all the baselines, we observe that FedPAQ has the longest computation time, which verifies that it takes more rounds to reach the same accuracy. Though FedPAQ reduces the communication time of each round aggressively, the increased number of training rounds makes the total training time longer than others.

\subsection{Different Levels of Non-IID Data}
\label{sec:eval_noniidness}

% Table 2:
% Cifar-100: The performance after same number of epeochs
% network, noniid-level, accuracy, data sent, wall-clock time

\begin{table}[t]
    \centering
    \resizebox{\columnwidth}{!}{
    \begin{tabular}{m{0.3cm}<{\centering} m{0.9cm}<{\centering} m{0.9cm}<{\centering} m{1.5cm}<{\centering} m{3.0cm}<{\centering} }
    \hline
    \hline
    $\sigma_d$ & Method  & Avg. rounds & Avg. data uploaded & Total time (Second)\\
    \hline
    \hline
    \multirow{6}{*}{0.2}
    & FedAvg  & 19.75 & 5.46 (1$\times$)  & 2538.56 $\pm$ 50.21 (1.31$\times$)\\ \cline{2-5}
    & QSGD  & 70.50 & 4.73 (1.15$\times$)  & 2144.46 $\pm$ 43.24 (1.56$\times$)\\ \cline{2-5}
    & Top-k  & 78.50 & 4.21 (1.30$\times$)  & 2187.28 $\pm$ 48.43 (1.53$\times$)\\ \cline{2-5}
    & FedPAQ  & 37.75 & 2.44 (2.24$\times$)  & 3338.05 $\pm$ 52.42 (1$\times$)\\ \cline{2-5}
    % & X  & Y & X  & Y \\ \cline{2-5}
    & \textbf{AdaGQ}  & \textbf{69.25} & \textbf{2.39 (2.29$\times$)}  & \textbf{1295.97 $\pm$ 40.10 (2.58$\times$)} \\ 
    \hline
    \hline
    \multirow{6}{*}{0.5}
    & FedAvg  & 23.00 & 6.36 (1$\times$) & 2919.34 $\pm$ 53.90 (1.35$\times$)\\ \cline{2-5}
    & QSGD  & 82.25 & 5.50 (1.16$\times$) &  2476.70 $\pm$ 46.32 (1.60$\times$)\\ \cline{2-5}
    & Top-k  & 85.50 & 4.61 (1.38$\times$) & 2411.62 $\pm$ 51.20 (1.64$\times$)\\ \cline{2-5}
    & FedPAQ  & 45.25 & 2.91 (2.19$\times$) & 3952.95 $\pm$ 58.23 (1$\times$)\\ \cline{2-5}
    % & X  & Y & X  & Y \\ \cline{2-5}
    & \textbf{AdaGQ}  & \textbf{82.75} & \textbf{2.86 (2.22$\times$)}  & \textbf{1558.92 $\pm$ 42.36 (2.54$\times$)} \\
        \hline
        \hline
    \multirow{6}{*}{0.8}
    & FedAvg  & 28.25 & 7.81 (1.26$\times$) & 3553.98 $\pm$ 57.30 (1.31$\times$)\\ \cline{2-5}
    & QSGD  & 146.50 & 9.83 (1$\times$) & 4409.73 $\pm$ 50.54 (1.06$\times$)\\ \cline{2-5}
    & Top-k  & 122.50 & 6.57 (1.50$\times$) & 3421.14 $\pm$ 54.32 (1.36$\times$)\\ \cline{2-5}
    & FedPAQ  & 52.75 & 3.41 (2.88$\times$) & 4655.70 $\pm$ 63.47 (1$\times$)\\ \cline{2-5}
    % & X  & Y & X  & Y \\ \cline{2-5}
    & \textbf{AdaGQ}  & \textbf{142.25} & \textbf{4.90 (2.00$\times$)}  & \textbf{2667.07 $\pm$ 48.92 (1.75$\times$)} \\    \hline \hline
    \end{tabular}
    }
    \caption{GoogLeNet on Cifar-10 under different $\sigma_d$}
    \label{tab:non-iid_level_googlenet}
    %\vspace{-0.3in}
\end{table}

In this section, we evaluate AdaGQ under different levels of non-iid data (i.e., different $\sigma_d$) against the baselines. 
% $\sigma_d = 1.0$ means that the data on one client only belongs to one class; 
% Specifically, 
% $\sigma_d = 0.8$ means that 80\% of the data on one client belongs to one class and the remaining 20\% of the data uniformly belongs to other classes;
% $\sigma_d = 0.5$ means that 50\% of the data on one client belongs to one class and the remaining 50\% of the data uniformly belongs to other classes;  $\sigma_d = 0.2$ means that 20\% of the data on one client belongs to one class and the remaining 80\% of the data uniformly belongs to other classes.
Table~\ref{tab:non-iid_level_resnet} and Table~\ref{tab:non-iid_level_googlenet} show the results by training ResNet-18 and GoogLeNet on Cifar-10, respectively. Owing to space limitation, we do not present the results on FMNIST, which share similar observations as those on Cifar-10. 
For each FL task, we evaluate the algorithms in terms of the total number of communication rounds, the average amount of uploaded data per client (in GB), and the total time (in second) to reach the same accuracy. We repeat the evaluation four times and report the average of those metrics.

From the tables we observe that AdaGQ outperforms all baseline algorithms in terms of total time under various non-iid levels. 
Among the baselines, FedAvg has the fewest communication rounds under all levels of non-iid data due to the periodic averaging to reduce the communication frequency. 
Here, we clarify that FedAvg (and also FedPAQ) has five epochs in each round, so the total number of epochs is five times the number of communication rounds. 
For example, in Table~\ref{tab:non-iid_level_resnet} when $\sigma_d=0.5$, FedAvg has about 71 ($14.25\times 5$) epochs which imply a longer computation time than QSGD ($\sim$48), Top-k ($\sim$46) and AdaGQ ($\sim$47).
Meanwhile, without any gradient compression, the amount of data transmitted each round in FedAvg is much higher than other algorithms, leading to longer communication time and thus longer total time.
In addition, we observe that FedPAQ has the most training epochs, which results in the longest computation time and a long total time.
An interesting observation is that, in Table~\ref{tab:non-iid_level_resnet} with $\sigma_d=0.8$, AdaGQ has more communication rounds than Top-k (68.0 v.s. 60.5) and 5.92\% less data uploaded, while achieving 28.8\% less total time.
Such a big improvement may be attributed to the heterogeneous quantization. Although AdaGQ has 5.40GB data uploaded on average per client, the slowest clients may have much less data to upload, which greatly reduces the communication overhead caused by waiting for the slowest clients in each round.

The non-iid level of data distribution affects the convergence
speed of training. A higher level of non-iid data decreases the convergence
speed in general, which results in more communication
rounds, thus more data uploaded and longer total
time. For example, the average number of communication rounds of AdaGQ
when training ResNet-18 on Cifar-10 with non-iid levels of 0.2, 0.5,
and 0.8 are 43.25, 47.00, and 68.00, respectively, which is increasing.
Similar conclusion is also suggested by other algorithms.

\subsection{Levels of Resource Heterogeneity}

In this section, we evaluate AdaGQ under different levels of resource heterogeneity.
To isolate the effects of resource heterogeneity, we fixed the dataset of each client with a non-iid level to be 0.5 for each running of the experiment.
We define the resource heterogeneity level $\sigma_r$ to be the ratio of the data transmission rate of the fastest client and that of the slowest client. 
We set the transmission rate of the fastest client to be 20Mbps, and the slowest client to be $20/\sigma_r$ Mbps, and the transmission rates of other clients are sampled randomly between [$20/\sigma_r$, 20] Mbps.
% For example, $\sigma_r = 2$ means the data transmission rate of the fastest client is twice that of the slowest client.
Similar to Section~\ref{sec:eval_noniidness}, we repeat the evaluation four times and report
the average of the metrics. We only present the results for training ResNet-18 on Cifar-10 
(in Table~\ref{tab:resource_hetero_resnet}), since other FL tasks share similar observations.

For FedAvg, QSGD, Top-k and FedPAQ, the resource heterogeneity does not affect the number of their communication rounds and the amount of uploaded data, and only changes the communication time of each round due to the delay of aggregation caused by the slowest client.
AdaGQ is able to adapt the number of quantization levels based on the clients' resources, thus reducing more total training time under higher resource heterogeneity. For example, when training ResNet-18 on Cifar-10, AdaGQ reduces the total time by 38.8\% compared to Top-k (the second best algorithm) when $\sigma_r=6$, which is higher than 25.9\% that is achieved when $\sigma_r=2$.

% Table 3:
% Cifar-10: The performance after same number of epeochs
% network, hetero-level, accuracy, data sent, wall-clock time
\begin{table}[t]
    \centering
    \resizebox{\columnwidth}{!}{
    \begin{tabular}{m{0.3cm}<{\centering} m{0.9cm}<{\centering} m{0.9cm}<{\centering} m{1.5cm}<{\centering} m{3.1cm}<{\centering} }
    \hline
    \hline
    $\sigma_r$ & Method  & Avg. rounds & Avg. data uploaded & Total time (Second)\\
    \hline
    \hline
    \multirow{6}{*}{2}
    & FedAvg  & 14.25 & 7.13 (1$\times$)  & 1742.20 $\pm$ 38.29 (1.11 $\times$)\\ \cline{2-5}
    & QSGD  & 45.75 & 5.49 (1.30$\times$) & 1488.89 $\pm$ 55.32 (1.30 $\times$)\\ \cline{2-5}
    & Top-k  & 46.00 & 4.47 (1.60$\times$) & 1232.43 $\pm$ 56.49 (1.57 $\times$)\\ \cline{2-5}
    & FedPAQ  & 22.50 & 2.75 (2.60$\times$) & 1938.03 $\pm$ 51.35 (1$\times$)\\ \cline{2-5}
    % & X  & Y & X  & Y \\ \cline{2-5}
    & \textbf{AdaGQ}  & \textbf{43.75} & \textbf{4.07} (1.75$\times$)  & \textbf{913.70 $\pm$ 47.24 (2.12$\times$)} \\ 
    \hline
    \hline
    \multirow{6}{*}{4}
    & FedAvg  & 14.50 & 7.26 (1$\times$)  & 2378.65$\pm$ 43.16  (1$\times$) \\ \cline{2-5}
    & QSGD  & 47.50 & 5.64 (1.29$\times$) & 2176.67 $\pm$ 66.50 (1.09$\times$)\\ \cline{2-5}
    & Top-k  & 45.75 & 4.38 (1.66$\times$) & 1763.90 $\pm$ 67.82 (1.35$\times$)\\ \cline{2-5}
    & FedPAQ  & 23.00 & 2.81 (2.58$\times$) & 2269.42$\pm$ 52.27 (1.05$\times$) \\ \cline{2-5}
    % & X  & Y & X  & Y \\ \cline{2-5}
    & \textbf{AdaGQ}  & \textbf{47.75} & \textbf{3.78 (1.92$\times$)} & \textbf{1134.63 $\pm$ 55.08 (2.10 $\times$)} \\
        \hline
        \hline
    \multirow{6}{*}{6}
    & FedAvg  & 14.75 & 7.38 (1$\times$)  & 2996.6 $\pm$ 48.23 (1$\times$)\\ \cline{2-5}
    & QSGD  & 48.00 & 5.84 (1.26$\times$) & 2889.67 $\pm$ 68.90 (1.04$\times$)\\ \cline{2-5}
    & Top-k  & 47.50 & 4.66 (1.58$\times$) & 2321.29 $\pm$ 66.35 (1.29$\times$) \\ \cline{2-5}
    & FedPAQ  & 23.25 & 2.75 (2.68$\times$) & 2514.09 $\pm$ 52.54 (1.19$\times$)\\ \cline{2-5}
    % & X  & Y & X  & Y \\ \cline{2-5}
    & \textbf{AdaGQ}  & \textbf{53.25} & \textbf{3.87 (1.91$\times$)} & \textbf{1419.71 $\pm$ 61.34 (2.11$\times$)}\\    \hline \hline
    \end{tabular}
    }
    \caption{The performance of Resnet-18 on Cifar-10 under different level of resource heterogeneity}
    \label{tab:resource_hetero_resnet}
 %   \vspace{-0.1in}
\end{table}

\section{Related Work}
\label{sec:related}

\textbf{Communication-efficient federated learning.} FL has been widely deployed for mobile and IoT devices. 
To reduce the communication bottleneck, various methods have been proposed which fall into two main categories. 
The first category reduces the communication overhead by periodic averaging which allows clients to perform multiple rounds of local updates and upload the updates less frequently \cite{mcmahan-aistat2017, wang-mlsys2019, zhao-icdcs2019, haddadpour-nips2019}. 
The second category of research solves this problem by reducing the communication overhead of every communication round \cite{wangni-nips2017, vogels-nips2019, han-icdcs2020, alistarh-nips2017, wen-nips2017, albasyoni2020optimal, ozfatura-isit2021, mao-TIST2022}. 
In this category, a variety of compression schemes have been proposed, including gradient quantization \cite{seide-interspeech2014, wen-nips2017, alistarh-nips2017}, gradient sparsification \cite{wangni-nips2017, han-icdcs2020, li-ggs2020} and low-rank approximation \cite{vogels-nips2019}.
Seide {\em et al.} \cite{seide-interspeech2014} replaced each weight with just the sign values.
Similarly, Wen {\em et al.} \cite{wen-nips2017} proposed TernGrad which requires three numerical levels \{−1, 0, 1\}, to aggressively reduce the communication time.
However, these two gradient quantization algorithms lack flexibility in controlling the resolution of quantization.
Alistarh {\em et al.} \cite{alistarh-nips2017} proposed quantized SGD (QSGD) that can adjust the number of bits (i.e., quantization resolution) sent per iteration to reduce the bandwidth cost, which provides more flexibility. However, 
how to find the optimal quantization resolution is not studied.
Han {\em et al.} \cite{han-icdcs2020} proposed an adaptive approach for gradient sparsification (i.e., Top-k) to achieve the near-optimal communication and computation trade-off by controlling the degree of gradient sparsity. Although they seek to find the optimal degree of gradient sparsity, the optimal value is assumed to be fixed. Besides, there are also some literature combines the two directions by integrating gradient quantization in periodic averaging \cite{reisizadeh-aistats2020}.
Different from them, we do not assume a fixed quantization resolution given the variations of gradient value during the training process.

% *** FL under heterogeneous clients
\textbf{Federated learning under heterogeneous clients.} Considering the heterogeneity of edge devices, FL under heterogeneous clients have also been studied in recent literature \cite{li-mobicom2021, wang-tsp2021, ghosh2019robust, sattler-tnls2019, wang-icdcs2019, diao-iclr2020, zhao-icdcs2019, ma-2021ijcai, li-infocom2021}.
Some of them consider the data heterogeneity across devices \cite{li-mobicom2021, wang-tsp2021, sattler-tnls2019}.  
For example, Li {\em et al.} \cite{li-mobicom2021} proposed a subnetwork based approach that aims to improve inference accuracy by learning personalized models. Though the proposed framework also reduces the communication cost, the heterogeneous communication resources are not considered and hence the clients with poor communication conditions can still be the bottleneck.
In \cite{wang-icdcs2019}, Wang {\em et al.} proposed an approach that identifies irrelevant updates of clients and precludes the uploading of these updates to save bandwidth.
Considering the resource heterogeneity, the asynchronous aggregation strategy has been designed to address the straggler problem \cite{zhao-icdcs2019, ma-2021ijcai}, where the server aggregation does not have to wait for all clients. Although the asynchronous aggregation reduces the delay by stragglers, the delayed gradients of stragglers introduce errors or even diverge the learning of the model.
Different from them, we propose heterogeneous gradient quantization to reduce the communication time of stragglers without compromising the model performance.

\section{Conclusions}
\label{sec:conclusion}

In this paper, we proposed AdaGQ, an adaptive and heterogeneous gradient quantization algorithm for communication-efficient federated learning for mobile edge devices.  
Based on varying gradient norm during training, we proposed an adaptive gradient quantization to seek the optimal quantization resolution in an online manner to minimize the total training time. 
We further designed heterogeneous gradient quantization to 
align the training time of slow clients in each round with others to mitigate the straggler effects. 
Evaluations based on various models and datasets validate the effectiveness of AdaGQ.

%\clearpage
%\vspace{-0.1in}

%------------------------------------------------------------------------- 
{
\footnotesize
% \balance
\bibliographystyle{IEEEtran}
\bibliography{main}
}

\end{document}